\documentclass[Afour,sagev,times,doublespace]{sagej}

\usepackage{moreverb,url}
\PassOptionsToPackage{hyphens}{url}

\usepackage{graphicx}
\usepackage{comment}
\usepackage{tabularx}
\usepackage{makecell}

\newcolumntype{Y}{>{\raggedright\arraybackslash}X}

\usepackage[none]{hyphenat}
\usepackage{amsfonts}
\usepackage{makecell}

\usepackage[utf8]{inputenc}
\usepackage{amsmath}

\usepackage[colorlinks,bookmarksopen,bookmarksnumbered,citecolor=red,urlcolor=red]{hyperref}

\newcommand\BibTeX{{\rmfamily B\kern-.05em \textsc{i\kern-.025em b}\kern-.08em
T\kern-.1667em\lower.7ex\hbox{E}\kern-.125emX}}

\def\volumeyear{2024}

\begin{document}

\runninghead{Nugent et al., TMLE for partially clustered trials}

\title{Causal Inference in Randomized Trials with Partial Clustering}

\author{Joshua R. Nugent\affilnum{1}, Elijah Kakande\affilnum{2}, Gabriel Chamie\affilnum{3}, Jane Kabami\affilnum{2}, Asiphas Owaraganise\affilnum{2}, Diane V. Havlir\affilnum{3}, Moses Kamya\affilnum{2,4}, Laura B. Balzer\affilnum{5}}

\affiliation{\affilnum{1}Division of Research, Kaiser Permanente Northern California, Pleasanton, CA, USA.\\
\affilnum{2}Infectious Diseases Research Collaboration, Kampala, Uganda.\\
\affilnum{3}Department of Medicine, University of California, San Francisco, CA, USA.\\
\affilnum{4}Department of Medicine, Makerere University, Kampala, Uganda.\\
\affilnum{5}Division of Biostatistics, School of Public Health, University of California, Berkeley, CA, USA.}
\corrauth{Joshua R. Nugent, Division of Research, Kaiser Permanente Northern California, 4480 Hacienda Drive, Pleasanton, CA 94588 USA.}
\email{joshua.r.nugent@kp.org}

\begin{abstract}
In some randomized trials, participants are not independent from another. The dependence, also referred to as ``clustering," may arise due to shared environmental factors, interactions between individuals, and/or the delivery of the intervention. For example, in some cluster randomized trials (CRTs), pre-existing clusters (such as neighborhoods or schools) are enrolled, randomized to the study conditions, and serve as the basis of delivery of the intervention or control. Such CRTs are considered to be ``fully clustered'': participants are dependent within clusters. 
In contrast, ``partially clustered'' trials contain a mix of participants that are dependent within clusters and participants that are completely independent of all others. One example of this design is a trial where participants are artificially grouped together for the purposes of randomization only; then, for intervention participants, the groups are the basis for intervention delivery, while control participants are un-grouped and continue with standard practice. Another example is an individually randomized group treatment trial (IRGTT) where participants are individually randomized and, post-randomization, intervention participants are grouped together for intervention delivery, while the control participants remain un-grouped and continue with with a standard practice or experience an alternate (ungrouped) intervention. For these three trial designs, we use causal models to non-parametrically describe the data generating process and formalize the dependence in the observed data distribution. In the process, we show that despite the different randomization approach, both partially clustered designs can be represented with the same dependence structure. Therefore, we can use the same statistical methods for estimation and inference of causal effects in these partially clustered trials. Specifically, we propose a novel implementation of targeted minimum loss-based estimation (TMLE) for partially clustered trials. TMLE is a model-robust approach, leverages covariate adjustment and machine learning to improve precision, and facilitates estimation of a large set of causal effects. In finite sample simulations, TMLE achieved comparable or had markedly higher statistical power than common alternatives for partially clustered designs. Finally, application of TMLE to real data from the SEARCH-IPT trial resulted in 20-57\% efficiency gains, demonstrating the real-world consequences of our proposed approach.
\end{abstract}


\maketitle

\onecolumn

\noindent November 2024 \\ Total word count: 324 abstract, 4758 main text \\Grant support: NIAID R01AI125000\\Trial registration number: NCT03315962


\section{Background}

In some randomized trials, participants are not independent from another. Dependence or ``clustering" may arise due to shared environmental factors, interactions between individuals, and/or intervention delivery. Consider a cluster randomized trial (CRT) where groups of participants are randomized and the resulting treatment conditions are applied to those groups. In many CRTs, the clusters are naturally occurring. For example, a trial evaluating teaching techniques might randomize classrooms, and within each classroom, student outcomes would be correlated due to having the same teacher, learning together, and the teaching technique implemented. In such CRTs, the typical assumption that each participant's outcome is independent of others' is violated,  compromising statistical inference if clustering is ignored \cite{donner_design_2010,hayes_cluster_2017}. These CRTs are considered to be ``fully clustered''\cite{lange_partially_2023}: participants are dependent within clusters, and the cluster is the unit of independence.  A variety of methods have been developed for analysis of fully clustered trials. For reviews, we refer the reader to \cite{turner_review_2017, benitez_defining_2023, murray_essential_2020}.

``Partially clustered trials"\cite{lange_partially_2023} refer to designs containing a mix of participants that are dependent within clusters and participants that are completely independent of all others. We refer the reader to Lange et al.\cite{lange_partially_2023} for a detailed overview of various trial designs resulting in partial clustering. We focus on two types of partially clustered trials.
First, we consider a trial where randomization occurs at the cluster-level, though some participants may be independent from all others; the SEARCH-IPT trial (NCT03315962), our motivating example, had this design\cite{kakande_mid-level_2022}. In SEARCH-IPT, groups of health managers were created and randomized to the study conditions: a mini-collaborative to improve tuberculosis prevention (treatment/intervention) or standard practice (control). After randomization, participants in treatment clusters received the group-level intervention, while participants in the control arm continued independently with the standard of care. As we detail below, in SEARCH-IPT, intervention participants were dependent within clusters, while control participants were effectively independent.

Second, we consider a partially clustered trial where randomization occurs at the individual level, but the intervention induces some clustering. Lange et al. classify these as ``individual randomization with post-randomization clustering"\cite{lange_partially_2023}. This is also referred to as an ``individually randomized group treatment trial" (IRGTT) or a ``partially nested" design\cite{thombs_effects_2021, pals_ignoring_2011, turner_review_2017, lange_partially_2023, moyer_evaluating_2024, christopher_mindfulness-based_2020}:  participants are randomized at the individual level, but a group-based treatment is delivered in at least one of the study conditions. For example, an IRGTT to improve mental health might involve group therapy sessions in one arm and individual care in the other arm. As we detail below, in such trials, intervention participants are dependent, while control participants are independent.

Figure~\ref{designs} provides a schematic of the trial designs discussed herein: a fully clustered CRT with randomization to pre-existing clusters and dependence within clusters in both arms; a partially clustered design with cluster-level randomization but intervention-only outcome dependence (e.g., SEARCH-IPT); and a partially clustered design with individual-level randomization and intervention-only outcome dependence (e.g., an IRGTT). 

Our goals in this paper are as follows. First, we use causal models to non-parametrically describe the data generating process in the three trial designs of interest. To the best of our knowledge, we are the first to formally study the dependence structure arising from the partially clustered designs. Second, we prove that while their designs differ, these  trials result in the same factorization of the joint distribution of the observed data; therefore, we can use the same approach for statistical estimation and inference. Next, we formally define causal effects that may be of interest in such trials \cite{benitez_defining_2023}. Importantly, these effects are defined through the causal models and, thus, agnostic to the algorithm used for estimation. Due to randomization, these effects are also easily identifiable (i.e., a function of the observed data distribution). Finally, for estimation and inference, we propose a novel implementation of targeted minimum loss-based estimation (TMLE), harnessing machine learning and accounting for the partially clustered structure \cite{van_der_laan_targeted_2011}. To the best of our knowledge, this is the first study of TMLE in trials with partial clustering, which have typically been analyzed using generalized linear mixed models\cite{bauer_evaluating_2008, candlish_appropriate_2018, roberts_design_2016} (GLMMs) or generalized estimating equations\cite{hibbs_accounting_2010, roberts_design_2016, wang_designing_2024} (GEEs). Using finite sample simulations, we demonstrate that TMLE  substantially improves statistical power when compared to these common alternatives. Finally, we illustrate the real-world consequences of our work with application to the SEARCH-IPT trial.

\subsection{Motivating study}
\label{motivating_study}

This work is motivated by the SEARCH-IPT trial, which aimed to increase uptake of isoniazid preventative therapy (IPT) to prevent tuberculosis among persons with HIV in Uganda (NCT03315962) \cite{kakande_mid-level_2022}. The intervention was tailored to district-level health managers, who oversee implementation of Ministry of Health guidelines in their respective districts. For the purposes of randomization, 82 health managers were grouped into 14 clusters of 4-7 managers. The groupings were based on the district's region, population density, and number of persons with HIV receiving care. Randomization resulted in 7 clusters per arm. Among managers randomized to the intervention, these groups became the mini-collaboratives in which a network-based intervention to improve IPT knowledge and uptake was delivered. In contrast, managers randomized to the control continued with standard practice and, as detailed below, are effectively independent. Full details of the study and results have been published separately \cite{kakande_mid-level_2022, balzer_statistical_2021}. Here, we use SEARCH-IPT as a motivating example to propose and evaluate a novel approach to effect estimation in partially clustered designs with cluster-level randomization and explicitly draw the connection to other designs with partial clustering, such as IRGTTs.

\section{Methods}
\subsection{Causal models representing the data generating process}
\label{causalmodel}

\begin{figure}[ht]
\begin{center}
\includegraphics[width=\textwidth]{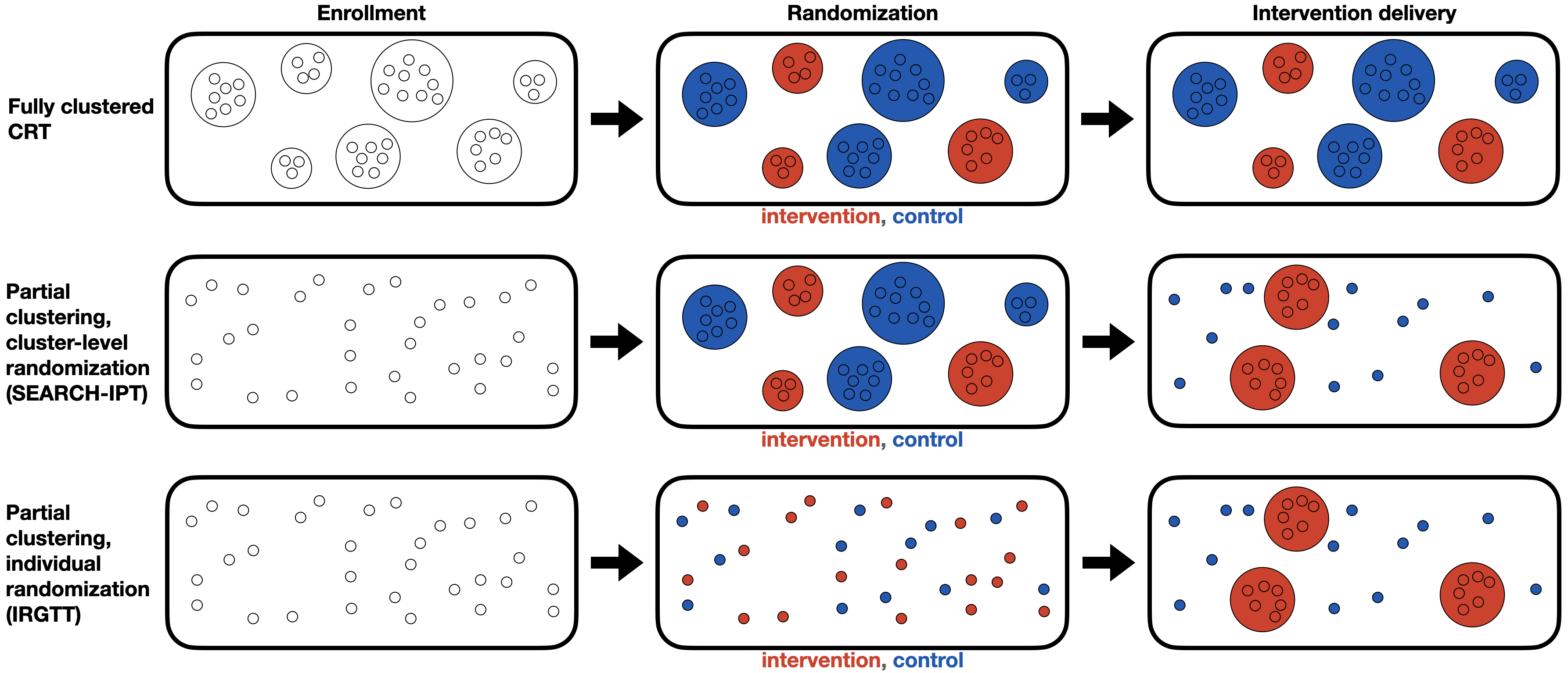}
    \caption{Schematic of different trial designs: (1) a fully clustered design with cluster-level randomization and dependence within clusters in both arms, (2) a partially clustered design with cluster-level randomization but intervention-only outcome dependence, such as SEARCH-IPT, and (3) a partially clustered design with individual randomization, post-randomization grouping, and intervention-only outcome dependence (e.g., an IRGTT). The small circles represent participants, and the larger circles represent clusters of participants.}
    \label{designs}
\end{center}
\end{figure}

In the following, we specify hierarchical causal models to formally represent the three trial designs shown in Figure~\ref{designs}. Throughout, we focus on trials with two arms, but our results naturally generalize to trials with more arms. Throughout, we use ``participants'' or ``individuals" to the describe the units that might be  grouped or ungrouped before or after randomization. 
We first introduce the notation and refer the reader to Appendix~\ref{app:notation} for a notation compendium. Throughout, individuals are indexed by $i$ and clusters by $j$. Un-bolded text (e.g., $W_{ij}, A_j$) is used for variables defined for a single individual or single cluster. Bold type with a subscript (e.g., $\mathbf{W}_j$) is used for variables  collected across multiple individuals in a cluster. Bold type with a superscript (e.g., $\mathbf{W}^J, \mathbf{A}^J$) is used for variables collected across multiple clusters in a trial.

\subsubsection{Fully clustered trial with cluster-level randomization:}
\label{CRT}
Here, we specify a hierarchical causal model to describe the data generating process for a fully clustered trial with naturally occurring clustering in both arms \cite{balzer_new_2019, benitez_defining_2023, balzer_two-stage_2021, nugent_blurring_2022}.

First, clusters $j=1,\ldots, J$ are defined and selected from the target population. Participants in each cluster are indexed by $i=1, \dots, N_j$. The cluster size $N_j$ may be constant, but more commonly varies.  Next, baseline covariates are measured. Let $W_{ij}$ denote the baseline covariate values of the $i^{th}$ participant in cluster $j$ and $\mathbf{W}_j=(W_{ij}:i =1,\ldots, N_j)$ be the corresponding set of covariates for all participants in cluster $j$. Usually, $W_{ij}$ and, thereby, $\textbf{W}_j$ include cluster-level characteristics. Clusters are then randomized to the intervention or control condition. Let $A_j=1$ indicate that cluster $j$ was randomized to the intervention; by design, all participants in a given cluster receive the same treatment. Finally, outcomes are measured among participants; the outcome vector  for cluster $j$ is denoted $\mathbf{Y}_j = (Y_{ij}: i = 1, \dots , N_j)$.

The data generating process for this CRT can be represented by drawing $J$ times from the following non-parametric structural equation model (NPSEM) \cite{pearl_causality_2009, balzer_new_2019}:
\begin{equation}
\label{eq:SCM}
\begin{aligned}
    \mathbf{W} & = f_{\mathbf{W}} (U_{\mathbf{W}})\\
    A &=f_{A}(U_{A})\\
    \mathbf{Y} & = f_{\mathbf{Y}}( \mathbf{W}, A, U_{\mathbf{Y}}).
\end{aligned}
\end{equation}
where $\textbf{U} = (U_\mathbf{W}, U_A,  U_{\mathbf{Y}})$ denote unmeasured factors contributing to the baseline covariates, intervention assignment, and outcomes, respectively. Usually, $(U_\mathbf{W},U_{\mathbf{Y}})$ are correlated between participants within a cluster. For example, environmental factors are often unmeasured, shared across participants, and influence their baseline covariates and outcomes. By design, however, the unmeasured factors contributing to the intervention assignment $U_{A}$ (i.e., randomization) are independent of the others. As detailed in \cite{balzer_new_2019, benitez_defining_2023}, this causal model accommodates many sources of dependence between participants within a cluster including heterogeneous levels of dependence by arm and by cluster. However, this model reflects that the $J$ clusters in the trial are independent. Letting $\textbf{O}=(\textbf{W},A,\textbf{Y})$ denote the observed data for a given cluster, the joint distribution of the observed data  $\textbf{O}^J = (\textbf{O}_1, \ldots, \textbf{O}_J)$ factorizes into a product across the clusters:  

\begin{equation}
\begin{aligned}
\mathbb{P}(\mathbf{O}^J)   = \prod_{j=1}^J \mathbb{P}(\mathbf{O}_j) = \prod_{j=1}^J\mathbb{P}(\textbf{W}_j)  \mathbb{P}(A_j)  
\mathbb{P}(\textbf{Y}_j \mid A_j, \textbf{W}_j).
\end{aligned}
\end{equation}

\subsubsection{Partially clustered designs - Overview:}
In the following subsections, we present causal models for the two partially clustered designs and study the resulting dependence structure. To facilitate comparison of these designs, we introduce the following notation. Let $N_T = \sum_j N_j$ denote the total number of participants. Following \cite{balzer_new_2019}, let $i\cdot= 1, \ldots, N_T$ index participants, pooling across clusters. For example, $Y_{i\cdot}$ denotes the outcome for the $i^{th}$ participant in the trial, and $W_{i\cdot}$ their covariates. Following \cite{van_der_laan_adaptive_2012, balzer_adaptive_2015}, let $\textbf{W}^J=(\textbf{W}_j:j =1,\ldots, J)$ denote the baseline covariates of all clusters and $\textbf{A}^J=(A_j: j=1, \ldots J$) denote the treatment assignments of all clusters. 
Finally, to differentiate between study arms and highlight the resulting dependence structure, we use $k = 1, \dots, K$ to index the $K$ intervention clusters of size $N_k$ and $l = 1, \dots, L$ to index the $L$ control ``clusters'' of size $N_l$.

\subsubsection{Partially clustered designs with cluster-level randomization (e.g., SEARCH-IPT):}
\label{not-irgtt}
We now consider a design resulting in partial clustering: participants are put into clusters; the clusters are randomized, participants in intervention clusters receive a group-based intervention, while participants in the control arm continue independently with standard practice (i.e., without an intervention). For this design, we  modify the above causal model to reflect (i) selecting all $N_T$ initially independent participants for inclusion in the trial; (ii) creating $J$ clusters as function of the baseline covariates of all participants; (iii) randomizing the study conditions to the resulting clusters; and (iv) measuring outcomes on participants.  The data generating process for this CRT design can be represented  by a single draw from the following NPSEM:

\begin{equation}
\label{eq:SCM2}
\begin{aligned}
    W_{i\cdot} &= f_{W} (U_{W_{i\cdot}}) \text{ for } i=1, \ldots, N_T \\
    \textbf{A}^J &=f_{\textbf{A}^J}(\textbf{W}^J, U_{\textbf{A}^J})\\
    \mathbf{Y}_j & = f_{\mathbf{Y}}( \mathbf{W}_j,  A_j, U_{\mathbf{Y}_j}) \text{ for } j=1, \ldots, J\\
    & \text{ where } \mathbf{Y}_k =  f_{\mathbf{Y}} (\mathbf{W}_k,  A_k, U_{\mathbf{Y}_k})
    \text{ for } k=1, \ldots, K \textbf{ if } A_j=1\\
    & \text{ and } Y_{il} =  f_{Y} (W_{il}, A_l, U_{Y_{il}})
    \text{ for } i=1, \ldots, N_l \text{ and } l=1, \ldots, L 
    \textbf{ if } A_j=0
\end{aligned}
\end{equation}
where the unmeasured factors contributing to the covariates $U_{W_{i\cdot}}$ and outcome  $U_{Y_{i\cdot}}$ are likely correlated within a participant, but independent from the unmeasured factors contributing to the cluster-level treatment assignment $U_{A_j}$. This causal model encodes that the baseline covariates of each participant are independently generated by some common function $f_W$. Additionally, this causal model reflects that by creating the clusters,  treatment assignment  vector  $\textbf{A}^J$ is a function of the covariates of all participants $\textbf{W}^J$.  As before, the outcomes in the $K$ intervention clusters  are correlated; they may depend on the characteristics of their cluster members ($\mathbf{W}_k)$, the shared exposure ($A_k=1$), as well as the joint error term ($U_{\textbf{Y}_k}$). In contrast, the outcomes in the $L$ control clusters are effectively independent. Specifically, $Y_{il}$, the outcome for participant $i$ in control cluster $l$ \emph{only} depends on their baseline covariates $W_{il}$, exposure $A_l=0$, and their specific unmeasured factors $U_{Y_{il}}$.
 
This causal model translates into the following set of assumptions on the joint distribution of the observed data: 

\begin{equation}
\begin{aligned}
\mathbb{P}(\mathbf{O}^J) 
&= \mathbb{P}( \textbf{W}^J) \times \mathbb{P}(\textbf{A}^J \mid \mathbf{W}^J ) \times   \prod_{j=1}^J  \mathbb{P}(\mathbf{Y}_j \mid A_j, \mathbf{W}_j)		 \\
& = c \times \underbrace{\left( \prod_{i=1}^{N_T}  \mathbb{P}(W_{i\cdot}) \right)}_{\text{Covariate distribution}} \times 
  \underbrace{\left( \prod_{k=1}^{K} \mathbb{P}(\mathbf{Y}_k \mid A_k=1, \mathbf{W}_k) \right)}_{\text{Intervention arm outcome dependence}} \times
     \underbrace{\left(  \prod_{l=1}^{L} \prod_{i=1}^{N_l} \mathbb{P}(Y_{il} \mid A_{l}=0, W_{il}) \right) }_{\text{Control arm outcome independence}}
\end{aligned}
\end{equation}
where we have replaced the treatment mechanism with a constant $c$. For example, in a two-armed trial with equal allocation probability, the treatment mechanism is $c=\prod_{j=1}^J \mathbb{P}(A_j=1 \mid \mathbf{W}^J )= 0.5^J$.
The key results are (i) the baseline covariate distribution factorizes as a product across the $N_T$ participants, (ii) the outcomes are dependent within intervention clusters, and (iii) the outcomes are effectively independent in the control arm. Therefore, the effective number of independent units is $J^*=K + \sum_l N_l$. In other words, despite cluster-level randomization, this design results in a partially clustered trial with dependence within intervention clusters and independence for control participants. As we discuss next, an IRGTT results in the same factorization of the observed data distribution, up to a constant, despite its distinct design.

\subsubsection{Partially clustered designs with individual randomization (e.g., IRGTT):}
\label{irgtt}

We  consider a second design resulting in partial clustering: participants are individually randomized and, post-randomization, intervention participants are put into clusters. 
The data generating process for an IRGTT is 
(i) selecting all $N_T$, otherwise independent, participants for inclusion in the trial, 
(ii) randomizing participants to the study conditions,
(iii) placing of intervention participants into clusters  for intervention delivery, 
and (iv) measuring outcomes on participants. For simplicity, we consider an IRGTT where the creation of intervention clusters is independent of the baseline covariates, but our results should generalize to more complex implementations.  As before, $i\cdot= 1, \ldots, N_T$ indexes participants across the trial, and $k=1,\ldots, K$ indexes intervention clusters. For parallel notation, we use $l=1,\ldots, L$ to index control ``clusters'' where, by design,  each has 1 participant: $N_l = 1, \forall l$. By design, we again have $K+L=J$ ``clusters''. The data generating process for an IRGTT can be represented  by a single draw from the following NPSEM:

\begin{equation}
\label{eq:scm_irgtt}
\begin{aligned}
    W_{i\cdot} &= f_{W} (U_{W_{i\cdot}}) \text{ for } i=1, \ldots, N_T \\
    A_{i\cdot} &= f_{A} (U_{A_{i\cdot}}) \text{ for } i=1, \ldots, N_T \\
    \mathbf{Y}_j & = f_{\mathbf{Y}}( \mathbf{W}_j,  A_j, U_{\mathbf{Y}_j}) \text{ for } j=1,\ldots, J \\
       & \text{ where } \mathbf{Y}_{k} =  f_{\mathbf{Y}} (\mathbf{W}_{k}, A_k, U_{\mathbf{Y}_k})
    \text{ for } k=1, \ldots, K \text{ if } A_j = 1\\
    & \text{ and } 
    Y_{il} =  f_{Y} (W_{il}, A_{il}, U_{Y_{il}}) \text{ for } i=1 \text{ and } l=1, \ldots, L \text{ if } A_{j}=0 .
\end{aligned}
\end{equation}
where for a given participant the unmeasured factors contributing to the covariates $U_{W_{i\cdot}}$ and outcome $U_{Y_{i\cdot}}$ are likely correlated, but independent from   the participant-level treatment assignment $U_{A_{i\cdot}}$.

As before, this causal model encodes the baseline covariates are independently generated by some common function $f_W$. Here, the randomization occurs for each participant separately. (We can consider alternative schemes such as pair-matching or stratification, but ignore these complications for simplicity.) Also as before, the  outcomes of participants randomized to the intervention are correlated; they may depend on the characteristics of their cluster members ($\mathbf{W}_k)$, the shared exposure $(A_k=1$), as well as the joint error term $U_{\mathbf{Y}_k}$. In contrast, the outcomes of participants randomized to the control \emph{only} depend on their specific baseline covariates, exposure, and unmeasured factors.

The IRGTT causal model translates into the following set of assumptions: 
\begin{equation}
\begin{aligned}
\mathbb{P}(\mathbf{O}^J) 	 
& = c' \times \left( \prod_{i=1}^{N_T}  \mathbb{P}(W_{i\cdot}) \right) \times 
  \left( \prod_{k=1}^{K} \mathbb{P}(\mathbf{Y}_k \mid A_k=1, \mathbf{W}_k) \right) \times
     \left(  \prod_{l=1}^{L} \prod_{i=1}^{1} \mathbb{P}(Y_{il} \mid A_{il}=0, W_{il}) \right)  \\
& = c' \times \underbrace{\left( \prod_{i=1}^{N_T}  \mathbb{P}(W_{i\cdot}) \right)}_{\text{Covariate distribution}} \times 
  \underbrace{\left( \prod_{k=1}^{K} \mathbb{P}(\mathbf{Y}_k \mid A_k=1, \mathbf{W}_k) \right)}_{\text{Intervention arm outcome dependence}} \times
     \underbrace{\left(  \prod_{l=1}^{L} \mathbb{P}(Y_{l} \mid A_{l}=0, W_{l}) \right) }_{\text{Control arm outcome independence}}
\end{aligned}
\end{equation}
where we have replaced the treatment mechanism with a constant $c'$. For example, in a two-armed trial with equal allocation probability, the treatment mechanism is $c'=\prod_{i=1}^{N_T} \mathbb{P}(A_{i\cdot}=1 ) =  0.5^{N_T}$.
Importantly, up until this constant, the factorized joint distribution of the observed data is the same in an IRGTT where intervention participants are grouped post-randomization and in a CRT where control participants are un-grouped post-randomization (Section~\ref{not-irgtt}). By design in both trials, we have $K$  independent \emph{clusters} in the intervention arm and $\sum_l N_l$ independent \emph{participants} in the control arm. The effective number of independent units in both designs is $J^*=K+\sum_l N_l$. As detailed below,  we can use the same methods to define and estimate causal effects in both types of trials, despite their distinct designs.

\subsection{Counterfactual outcomes, causal effects, and identification}
\label{targparams}

For each trial design, we define counterfactual outcomes through hypothetical interventions on the corresponding causal model. Specifically, we replace the treatment assignment mechanism  with a constant $a \in \{0,1\}$ \cite{pearl_causality_2009, van_der_laan_targeted_2011, balzer_new_2019}. The resulting vector $\mathbf{Y}_j(a)=(Y_{ij}(a): i=1,\ldots,N_j)$ is the set of participant-level outcomes that would be observed if, possibly contrary-to-fact, cluster $j$ were to receive intervention level $A_j = a$. As before, $Y_{i\cdot}(a)$ denotes the counterfactual outcome for the $i^{th}$  participant in the trial.

Causal effects are defined through summary measures of these counterfactual outcomes. In trials with full or partial clustering, we may be interested in ``cluster-level'' effects that weight each cluster (regardless of its size) equally, ``individual-level'' effects that weight each participant equally, or effects defined through an alternative weighting scheme \cite{benitez_defining_2023, su_model-assisted_2021, kahan_estimands_2023}. For simplicity, we focus our presentation on ``individual-level'' effects, and note our findings are applicable to other targets if the appropriate weights are applied \cite{benitez_defining_2023}. In SEARCH-IPT, we focused on the sample average treatment effect (SATE), which is a contrast of the empirical means of the counterfactual outcomes for the study participants:  $\psi(a)=\frac{1}{N_T} \sum_{i=1}^{N_T} Y_{i.}(a)$ for $a=\{0,1\}$  \cite{splawa-neyman_application_1990, imbens_nonparametric_2004, imai_essential_2009, balzer_targeted_2016}. A more common target is the population average treatment effect (PATE), which implicitly generalizes the trial impact to the target population of interest. For notational convenience, we drop the subscripts and define the PATE as $\mathbb{E} [Y(1)] - \mathbb{E} [Y(0)]$  on the difference scale and as $\mathbb{E}[{Y}(1)] \div \mathbb{E}[{Y}(0)]$ on the relative scale. 

Due to the randomization, identification of these effects is trivial. As reflected in the causal models (Eqs.~\ref{eq:SCM}, \ref{eq:SCM2}, and \ref{eq:scm_irgtt}), the unmeasured factors contributing to the intervention assignments are independent of the unmeasured factors contributing to the baseline covariates and outcomes. In other words, there is no unmeasured confounding and, by design, there is a positivity probability of receiving the intervention or control condition (i.e., the positivity assumption holds). Therefore, regardless of the trial design, we do not need to adjust for baseline covariates to isolate the causal effect of interest. In other words, causal effects can be estimated with a simple contrast in the average outcomes between arms (i.e., an unadjusted effect estimator). However, as detailed next, adjustment for covariates (including baseline outcome level) that are prognostic of the outcome can substantially improve precision \cite{fisher_statistical_1932, hayes_cluster_2017, tsiatis_covariate_2008, van_der_laan_targeted_2011, benitez_defining_2023,  balzer_adaptive_2016, balzer_adaptive_2024, teerenstra_sample_2023}.

\subsection{Statistical estimation and inference}
\label{est_inf}

Trial analyses usually incorporate covariate adjustment by specifying an outcome regression, characterizing how the expected outcome is related to the treatment and  covariates. 
In clustered data settings, GLMMs and GEEs are common approaches, employing a parametric regression model to derive point estimates and inference based on the coefficient for the treatment arm \cite{fitzmaurice_applied_2012, turner_review_2017, hayes_cluster_2017}. To account for outcome correlations, GLMMs use random intercepts (typically assumed to be normally distributed) for cluster membership, and GEEs use a working correlation matrix. By explicitly modeling the variance-covariance structure, GLMMs and GEEs naturally account for the  partially clustering in these trials. Indeed, approaches for estimation and inference for IRGTTs build on GLMMs (e.g., \cite{roberts_design_2005, lee_use_2005, bauer_evaluating_2008, candlish_appropriate_2018}) and GEEs (e.g., \cite{roberts_design_2013}). Previous work has highlighted the potential inferential challenges of these approaches (with and without finite sample corrections) when there are few independent units \cite{pan_small-sample_2002, kreft_introducing_1998, leyrat_cluster_2018, kahan_increased_2016, benitez_defining_2023}. Additionally, the applications of GLMMs and GEEs have tended to allow the form of the outcome regression determine the effect estimated \cite{hubbard_gee_2010, benitez_defining_2023}. For example, for binary outcomes, these approaches commonly employ the logit link function and, thereby, yield estimates of marginal odds ratios without covariate adjustment and conditional odds ratios with covariate adjustment. It is worth nothing, however, that recent advances have expanded the flexibility of GLMMs and GEEs \cite{wang_designing_2024}. 

Under many trial designs and dependence structures, TMLE can robustly estimate a variety of causal effects (see, e.g., \cite{van_der_laan_targeted_2011,  balzer_adaptive_2016, benitez_defining_2023, balzer_adaptive_2024, nugent_blurring_2022}), including those defined in Section~\ref{targparams}. In general, TMLE is a framework for constructing asymptotically linear substitution estimators of target effects. For the estimands of interest here, TMLE utilizes estimates of both the outcome regression and the propensity score (the conditional probability of being randomized to the intervention, given covariates). Precision is gained with improved estimation of these nuisance parameters. Indeed, TMLE is semi-parametric efficient and will achieve the smallest asymptotic variance under consistent estimation of the outcome regression and propensity score. Therefore, for nuisance parameter estimation, we recommend Super Learner, an ensemble method employing sample-splitting to combine fits from a diverse set of candidate learners  \cite{van_der_laan_super_2007}. Additionally, a more nuanced implementation of TMLE harnesses Adaptive Pre-specification to flexibly and automatically select the adjustment approach maximizing empirical efficiency in randomized trials \cite{balzer_adaptive_2016, balzer_adaptive_2024}. 

To account for partial clustering in the trial designs of interest, we need to make two adjustments to the standard TMLE implementation (Appendix~\ref{app:tmle}). First, any sample-splitting or cross-validation scheme must reflect the dependence structure\cite{van_der_laan_targeted_2011, phillips_practical_2023}. 
Practically, for trials where control participants are un-grouped after randomization (e.g., SEARCH-IPT) or trials where intervention participants are grouped together after randomization (i.e., IRGTTs), all participants from a given intervention cluster will be assigned to the same split, while control participants will be randomly assigned to splits. Second,  statistical inference must reflect the dependence structure. By design, TMLE solves the efficient influence curve estimating equation for the effect of interest. Thus, a TMLE for the individual-level effect, which has been our focus, solves $0 = \frac{1}{N_T} \sum_{i=1}^{N_T} \hat{IC}_{i.}$ where $\hat{IC}_{i.}$ denotes the estimated influence curve (IC) for participant $i$. Following \cite{schnitzer_effect_2014, benitez_defining_2023} with details in   Appendix~\ref{app:ICagg}, we aggregate the participant-level IC to the cluster-level and estimate the variance of the TMLE with the sample variance of the aggregated ICs, divided by the number of independent units $J^*$. With this variance estimate, we obtain Wald-type 95\% confidence intervals and tests of the null hypothesis. Aggregation of the influence curve estimates in this manner does not require estimation or assumption of a within-cluster covariance structure\cite{van_der_laan_targeted_2018}. This TMLE can easily be implemented with the \texttt{ltmle} R package by specifying cluster membership with the \texttt{id} argument.

\subsection{Simulation study design}
\label{simstudy-description}

To evaluate estimator performance in trials  with partial clustering, we conducted a simulation study with correlated outcomes in the intervention arm but independent outcomes in the control arm. We explored 6 settings: $K \in \{30, 50\}$ intervention clusters with $N_k \in \{5,10,20\}$ participants. For each, the corresponding number of control ``clusters'' of size 1 was $L=K\times N_k$, giving us equal numbers of participants in each arm. For comparability across estimators, we  focused on a continuous outcome and the following data generating process:

\begin{itemize}
    \item Simulating the baseline covariates $W0_{i.} \sim \text{Normal}(0, 1)$ and  $W1_{i.} \sim \text{Bernoulli}(.5)$
    \item Assigning the intervention indicator $A_{i.} \sim \text{Bernoulli}(.5)$
    \item Placing intervention participants into $K$ clusters of size $N_k$; keeping the control participants in $L$ ``clusters'' of size 1
    \item Generating the outcome for participant  $i=\{1,\ldots, N_j\}$ in cluster $j=\{1,\ldots,J\}$ under 3 scenarios (n.b.: $J = K + L$):
	\begin{enumerate}
   	 \item \emph{Complex:} $Y_{i\cdot} =  \beta \cdot A_{i\cdot} + W0_{i\cdot}^2 - .5 \cdot W1_{i \cdot} + W0_{i\cdot}^2 \cdot W1_{i\cdot} + UY_{i\cdot} + UE_j$
     \item \emph{Main terms:} $Y_{i\cdot} =  \beta \cdot A_{i\cdot} + W0_{i\cdot} + .5 \cdot W1_{i\cdot} + UY_{i\cdot} + UE_j $
        	 \item \emph{Treatment only:} $Y_{i\cdot} =  \beta \cdot A_{i\cdot} + UY_{i\cdot} + UE_j$
\end{enumerate}
where the treatment coefficient $\beta$ was 0.25 when there was  an effect and 0 under the null. The unmeasured, participant-level factors were generated as $UY_{i.} \sim \text{Normal}(0,1)$. Within the intervention arm, we explored ``weaker'' and ``stronger'' levels of clustering by drawing an unmeasured, cluster-level factor $UE_{j} \sim \text{Normal}(0, \sigma^2)$ with $\sigma^2 \in \{0.03^2, 0.1^2, 0.23^2, 0.33^2\}$, corresponding to intracluster correlation coefficients in the intervention arm of .001, .01, .05 and .10 respectively. In the control arm, we set $UE_j=0$. 
\end{itemize}

Our estimand was the individual-level effect. For statistical estimation and inference compared:
\begin{enumerate}
    \item GLMM with identity link function, adjusting for ($W0,W1)$ as main terms,  including a random intercept  for any clusters of size $> 1$, and obtaining inference with the $t$-test with degrees of freedom determined by \cite{nugent_type_2021}.
     \item GEE with the identity link function, adjusting for ($W0,W1$) as main terms, using an exchangeable working correlation structure, and obtaining statistical inference via robust standard errors.
    \item TMLE with Super Learner to estimate the outcome regression and propensity score, with inference via the aggregated influence curve. Within Super Learner, we employed cross-validation to combine estimates from the simple mean, a main terms regression, and multivariate adaptive regression splines. The number of cross-validation splits was determined following recommendations in \cite{phillips_practical_2023} based on the number of independent units $J^*$.
\end{enumerate}

By design, the population average treatment effect $\mathbb{E}[Y(1)] - \mathbb{E}[Y(0)]$ was 0.25 when there was an effect and 0 under the null. From 2000 simulation iterations, we calculated bias as the average difference between the point estimate and true value, coverage as the proportion of 95\% confidence intervals that contained the true value, mean squared error (MSE) as $\text{bias}^2$ + variance, Type I error rate (when $\beta = 0$), and attained power (when $\beta \neq 0$). Since simulation studies include some random error,  we also calculated the Monte Carlo 95\% confidence intervals for these summary statistics.  

Additional simulations are described in more detail in Appendix~\ref{app:moresims}, including those with: i) similar parameters as described above, but with binary outcomes; ii) an unbalanced number of participants in the treatment and control arms; and iii) a version implementing TMLE without influence curve aggregation for inference.


\section{Results}
\subsection{Simulation study results}
\label{simstudy-results}

We expected all algorithms to perform well in the ``Main terms'' and ``Treatment only'' scenarios where GLMM and GEE were correctly specified and Super Learner (used within  TMLE) included the correctly specified outcome regression in its library. 
Indeed, all estimators displayed minimal bias, close to nominal coverage and Type I error, and had similar power and MSE (Appendix~\ref{app:moresims_cont}). GEE and TMLE approaches performed almost identically, with slight anti-conservative coverage compared to GLMMs. 

\begin{figure}
    \centering
    \includegraphics[width = \textwidth]{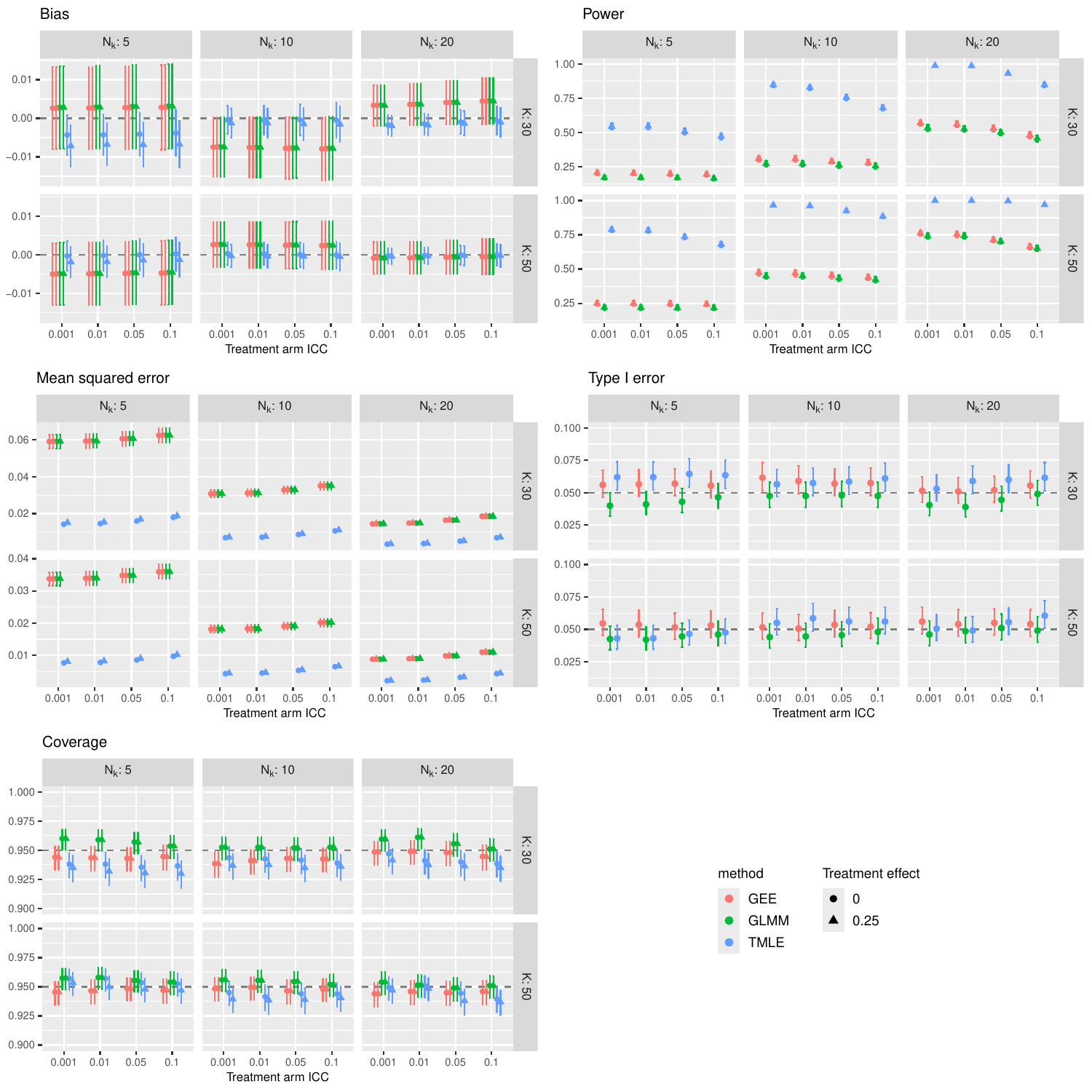}
    \caption{With 2000 iterations, simulation results for a continuous outcome under the ``Complex" data generating process, with and without an effect, under the following scenarios: $K=\{30,50\}$ intervention clusters each with $N_k=\{5,10\}$ participants and weaker-to-stronger clustering (i.e., the standard deviation of the random intercept) clustering. Each trial has $K\times N_k$ control ``clusters'' of size 1. The points represent the mean and error bars the 95\% Monte Carlo confidence intervals.}
    \label{fig:complex_covars}
\end{figure}

Conversely, in the ``Complex'' scenario, we expected TMLE to outperform the others, given its use of Super Learner for more flexible estimation.
As shown in Figure~\ref{fig:complex_covars}, estimator performance was markedly different under this setting. Across all sample sizes, TMLE achieved notably lower MSE and higher statistical power. For example, with $K=50$ intervention clusters of size $N_k=5$ and a treatment arm ICC of .05, TMLE achieved 3 times higher statistical power (73.6\%) than GEEs (24.6\%) or GLMMs (21.7\%). One tradeoff to note is that TMLE was slightly anti-conservative in this scenario, with mean coverage of 94.0\% under a non-null treatment effect and mean a Type I error rate of .056, both of which were more pronounced under the smallest sample sizes. In these smaller sample sizes, GEEs were also slightly anti-conservative, while GLMMs tended to be slighly conservative. All estimators displayed minimal bias.

In additional simulations, we found similar results for binary outcomes, though the gains in power from TMLE were slightly less dramatic (Appendix~\ref{app:moresims_bin}). Trials with an unbalanced number of participants by arm also performed similarly with both continuous (Appendix~\ref{app:moresims_cont_unbalanced}) and binary (Appendix~\ref{app:moresims_bin_unbalanced}) outcomes. Finally, as expected, implementing TMLE without aggregating the influence curve estimates in clusters led to significantly higher Type I error and lower coverage than the approach we proposed above (Appendix~\ref{app:sims_no_agg}). Code to reproduce the simulation studies is provided at \verb|https://github.com/joshua-nugent/SEARCH-IPT|. Simulations were conducted using R 4.3.1 with the \verb|ltmle|, \verb|SuperLearner|, \verb|gee|, \verb|lme4|, and \verb|lmerTest| packages.

\subsection{Application to SEARCH-IPT}
\label{sipt_results}

Recall that in SEARCH-IPT, district health officers were placed into groups for randomization \cite{kakande_mid-level_2022}. The groups randomized to the intervention became mini-collaboratives for a network-based strategy to improve IPT uptake. Participants randomized to the control arm continued independently with standard practice. Of the 82 participants, each corresponding to a health district, there were 7 intervention clusters of size 5-7 and 39 control ``clusters'' of size 1. The primary endpoint was the incidence rate of IPT initiation among persons with HIV in the district overseen by the participant and compared on the relative scale (i.e., incidence rate ratio) weighting districts equally; equivalent to the individual-level effect. The primary analysis employed TMLE with Adaptive Pre-specification \cite{balzer_adaptive_2016, balzer_adaptive_2024} to flexibly and automatically select the adjustment variable to maximize efficiency, while respecting the partial clustering\cite{balzer_statistical_2021}. In the primary analysis, this approach selected IPT uptake in the quarter prior to baseline as the sole adjustment variable.

The results of the SEARCH-IPT study have previously been published \cite{kakande_mid-level_2022}. The incidence rate ratio  was 1.27, with a log-scale standard error of  0.119 and a one-sided p-value of .026. In post-hoc analyses, we compared these results to GEE and GLMM. For the GEE, we modeled expected counts of the outcome as log-link Poisson with person-time-at-risk as an offset and assumed an exchangeable working correlation matrix with robust standard errors. For the GLMM, we used the log-link Poisson specification with a random intercept for the treatment arm clusters. Both models included a treatment effect indicator and the IPT uptake adjustment variable to match TMLE. For the GEE, the resulting rate ratio was 1.068 with a log-scale standard error of 0.117 and one-sided p-value of 0.29. For the GLMM, the resulting rate ratio of 1.073 with a log-scale standard error of 0.079 and a one-sided p-value of 0.18. Notably, these 3 estimators are targeting slightly different estimands, highlighting the importance of specifying the causal effect of interest and then conducting estimation and inference for that effect.

To demonstrate the importance of studying and respecting the dependence structure in partially clustered trials, we reanalyzed SEARCH-IPT under the following assumptions: partial clustering (i.e., the actual design), fully clustered (i.e., retaining the original control groupings), and fully unclustered (i.e., naively treating all units as independent). For each, we implemented TMLE adjusting for the IPT uptake in the quarter prior to baseline and obtained IC-based inference with the Student's $t$-distribution with $J^*-2$ degrees of freedom. For each implementation, we calculated the point estimate, variance estimate, efficiency gain as the ratio of the estimated variance of a given approach to the estimated variance with partial clustering, 95\% confidence interval, and p-value from the one-sided test of the null hypothesis that the SEARCH-IPT intervention did not improve uptake. 

The results are summarized in Table~\ref{applied_table_tmle}, overall and by the sex of patients in the health district. As expected by theory and our simulations, the point estimates were the same, but the variance estimates differed by clustering approach. The primary analysis that appropriately accounted for the partial clustering resulted in efficiency gains, overall and by subgroup. As compared to the fully clustered approach that assumes the randomization unit is the independent unit, the partially clustered TMLE was $\approx$ 20\% more efficient. As compared to the fully unclustered approach, the partially clustered TMLE was 40-57\% more efficient, highlighting the common misconception that clustered analyses are always less efficient than analyses ignoring clustering. These gains in precision were reflected in the corresponding 95\% confidence intervals and hypothesis tests, which additionally were impacted by the degrees of freedom (derived from the number of independent units).

\section{Conclusions}

Motivated by SEARCH-IPT, we  studied the theoretical and practical consequences of partially clustered trials for estimation and inference of causal effects. To the best of the knowledge, we are the first to use causal models to represent the data generating processes for a CRT where the control participants continue with the standard-of-care post-randomization and for an IRGTT where the intervention clusters are formed post-randomization. Also to the best of our knowledge, we are the first to formally show these designs lead to the same independence structure. In simulations, our novel implementation of TMLE resulted in markedly higher power in the ``Complex'' setting, which reflected the common scenario where prognostic covariates are known, but functional form is not. Implementation of the methods resulted in substantially different results compared to alternative approaches when applied to the analysis of SEARCH-IPT. We hope our presentation encourage researchers to use more flexible estimation approaches such as TMLE to evaluate intervention effects in trials with complex dependence. 

There are several limitations and avenues of future work. First, we focused on estimation and inference for individual-level effects, weighting participants equally. We plan to extend the weighting schemes in \cite{benitez_defining_2023} to estimate other effects with TMLE in partially clustered trials. Second, our simulation study was limited to designs with $K=\{30,50\}$ intervention clusters. 
With fewer clusters in the intervention or individuals in the control arm, the asymptotic guarantees for all  methods may not hold \cite{kahan_increased_2016, kauermann_note_2001}, and finite sample implementations to address this, such as TMLE with Adaptive Pre-specification \cite{balzer_adaptive_2016, balzer_adaptive_2024} are an area of future work. Third, we plan to formally study and compare the (in)dependence structure arising from other trials with partial clustering\cite{lange_partially_2023}, including trials with different levels of clustering in the two or more arms. Importantly, TMLE does not make assumptions about the covariance structure or have restrictions on cluster sizes. Fourth, we plan to generalize our results to trials where naturally occurring clusters (e.g., clinics) are randomized, but it reasonable to assume outcomes in the control arm are effectively independent. Likewise, we aim to generalize our results to observational studies where partial clustering is induced by the roll-out of cluster-level programs and policies. For both  settings, we anticipate  accounting for partial clustering will improve efficiency, but will also require nuanced approaches for confounding control \cite{balzer_new_2019, nugent_blurring_2022}. Finally, this work has consequences for meta-analyses of trials with the same intervention but different implementations (e.g., individual-level versus cluster-level randomization and delivery).


\clearpage
\newpage

\bibliographystyle{SageV}
\bibliography{sipt_sage}

\begin{acks}
On behalf of the SEARCH-IPT team, we thank the participating district health officers and district tuberculosis and leprosy supervisors for their generous participation in our trial, as well as the Uganda Ministry of Health.
\end{acks}

\begin{funding}
This work was supported, in part, by a grant from the National Institute of Allergy and Infectious Diseases (NIAID R01AI125000 [PI: DVH]). The funder (NIAID) had no role in the writing of the manuscript or decision to submit it for publication.
\end{funding}

\begin{dci}
None.
\end{dci}

\clearpage
\newpage

\clearpage
\newpage

\clearpage
\newpage

\begin{table}
\centering
\begin{tabular}{l|c|cccccc} 
        \thead{Group} & 
        \thead{Clustering \\ Approach}  & 
        \thead{Effective \\ Sample Size} & \thead{Point \\Estimate} & \thead{Variance \\ (Log-scale)} & \thead{Variance \\ Ratio} & \thead{95 \% \\ CI} & \thead{ \textit{p}-value\\ (one-sided) }\\
        \hline
        &Partial clustering & 46  &1.27 &.0141& 1     &1.00 - 1.61 &.026\\
        Overall  &Fully clustered& 14  & 1.27 &.0170& 1.20 &0.95 - 1.68 &.047\\
        &Fully unclustered        & 82  &1.27 &.0197& 1.40 &0.96 - 1.68 &.047\\
        \hline
        &Partial clustering & 46       &1.21 &.0155& 1   &0.94 - 1.55 &.067\\
        Women&Fully clustered  & 14    &1.21 &.0186& 1.20  &0.90 - 1.63 &.095\\
        &Fully unclustered    & 82      &1.21 &.0217& 1.40  &0.90 - 1.62 &.101\\
        \hline
        &Partial clustering  & 46 &1.27      &.0106&  1 &1.00 - 1.56 &.012\\
        Men &Fully clustered  & 14     &1.27 &.0129&  1.22&0.99 - 1.63 &.028\\
        &Fully unclustered      & 82    &1.27 &.0166&  1.57&0.98 - 1.64 &.033\\
\end{tabular}
\caption{Impact of assumed dependence structure on a re-analysis of the SEARCH-IPT trial.}
\label{applied_table_tmle}
\end{table}

\clearpage
\newpage

    
    
    

\appendix


\section{Notation compendium}
\label{app:notation}

Across this section, individuals are indexed by $i$ and clusters by $j$. Un-bolded text (e.g., $W_{ij}, A_j$) is used for variables defined for a single individual or single cluster. Bold type with a subscript (e.g., $\mathbf{W}_j$) is used for variables  collected across multiple individuals in a cluster. Bold text with a superscript (e.g., $\mathbf{W}^J, \mathbf{A}^J$) is used for variables collected across multiple clusters in a trial. 

\subsection{Fully clustered trial with cluster-level  randomization}
\label{notation1}

\begin{itemize}
    \item $J$: Number of clusters, indexed $j = 1, \dots, J$
    \item $N_j$: Number of individuals in cluster $j$, indexed $i = 1, \dots, N_j$
    \item $W_{ij}$: Baseline covariates of the $i^{th}$ individual in cluster $j$. May include cluster-level covariates
    \item $\mathbf{W}_j = (W_{ij}: i =1,\ldots, N_j)$: Set of covariates for all individuals in cluster $j$; hereafter ``covariate matrix'' for cluster $j$
    \item $A_j$: Indicator of cluster $j$'s treatment assignment to the intervention arm
    \item $\mathbf{Y}_j = (Y_{ij}: i = 1, \dots , N_j)$: Outcome vector  for cluster $j$
    \item $\textbf{O}=(\textbf{W},A,\textbf{Y})$: Observed data for a given cluster
    \item $\textbf{O}^J = (\textbf{O}_1, \ldots, \textbf{O}_J)$: Joint distribution of the observed data
\end{itemize}

\subsection{Partially clustered designs with cluster-level randomization (SEARCH-IPT)}
\label{notation2}

If not specifically noted here, all notation matches Appendix~\ref{notation1} above and extends it (rather than replacing it). Of note is that the ``$i\cdot$" below differs from the cluster-specific $i$ in earlier examples.

\begin{itemize}
    \item $N_T$: Total number of participants in the trial $(\sum_j N_j)$
    \item $W_{i\cdot}$: Covariates for the $i^{th}$ participant in the trial for $i=1,\dots,N_T$
    \item $Y_{i\cdot}$: Outcome for the $i^{th}$ participant in the trial for $i=1,\dots,N_T$
    \item $\textbf{W}^J=(\textbf{W}_j:j =1,\ldots, J)$: Baseline covariates of all clusters 
    \item $\textbf{A}^J=(A_j: j=1, \ldots J$): Treatment assignments of all clusters
    \item $K$: Number of clusters randomized to the intervention, indexed $k = 1, \dots, K$
    \item $L$: Number of  clusters randomized to the control, indexed $l = 1, \dots, L$
    \item $\textbf{W}_k$: Covariate matrix for intervention cluster $k$
    \item $\textbf{W}_l$: Covariate matrix for control cluster $l$
    \item $A_k = 1$, $A_l = 0$: Cluster-level treatment assignments for cluster $k$ in the intervention arm and cluster $l$ in the control arm
    \item $N_l$: Number of participants in control cluster $l$
    \item $J^*$: Number of independent units under this design, equal to $K + \sum_l N_l$
\end{itemize}

\subsection{Partially clustered designs with individual randomization (IRGTT)}
\label{notation3}

If not specifically noted here, all notation matches Appendix~\ref{notation1} and Appendix~\ref{notation2} above and extends it (rather than replacing it).

\begin{itemize}
    \item $A_{i\cdot}$: Treatment assignment for the $i^{th}$ participant in the trial for $i=1,\dots,N_T$
    \item $A_{il} = A_l = 0$: Treatment assignment (0) for participant $i = 1$ in singleton control ``cluster" $l$
    \item $W_{il} = W_l$: covariates for participant $i = 1$ in singleton control ``cluster" $l$
\end{itemize}


\section{Step-by-step TMLE implementation}
\label{app:tmle}

\begin{figure}
    \centering
    \includegraphics[width = \textwidth]{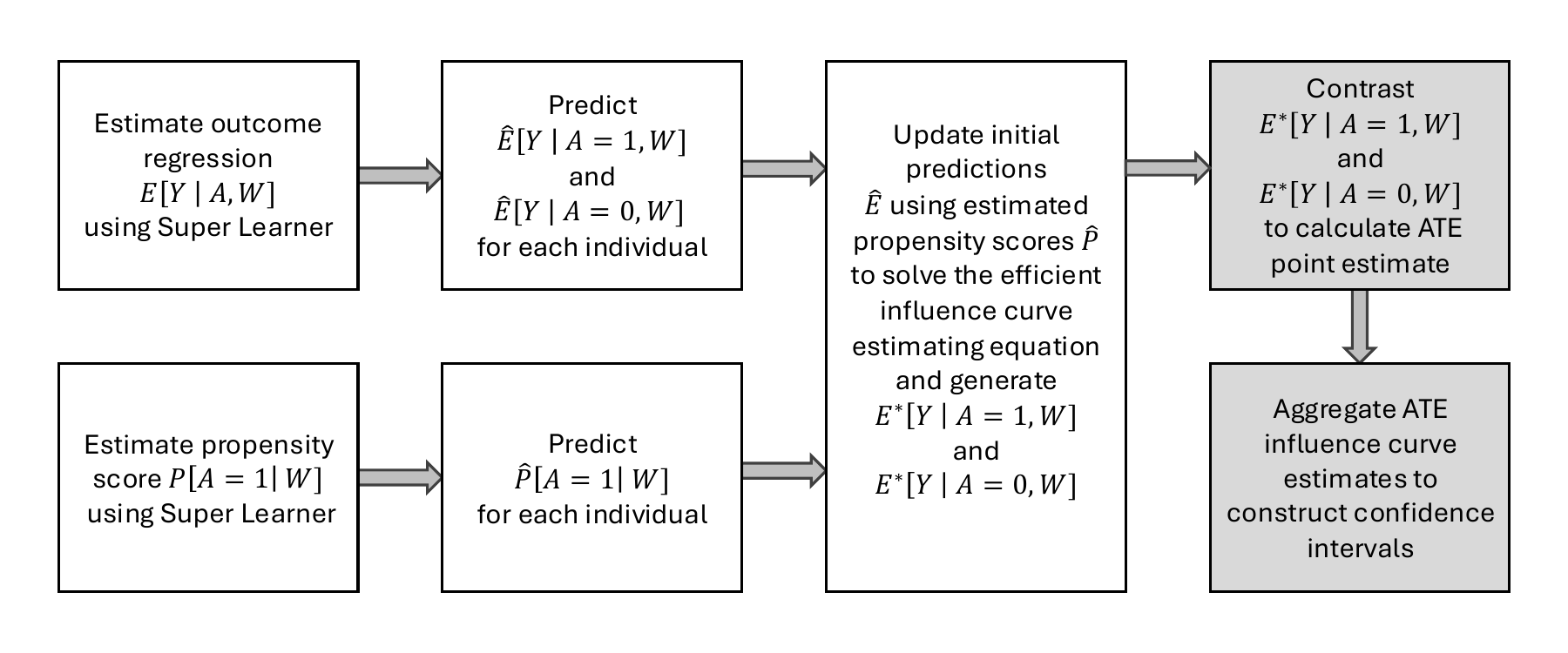}
    \caption{Schematic of TMLE for an individual-level effect with clustered or partially clustered data.}
    \label{fig:tmle_flowchart}
\end{figure}

We refer the reader to \cite{benitez_defining_2023} for a detailed presentation of TMLE to improve efficiency in trials with clustering. Here, we briefly review one implementation for estimating individual-level effects, where each participant is weighed equally (regardless of the size of their cluster).  Recall $(W_{i.}, A_{i.}, Y_{i.})$ are the baseline covariates, randomized treatment, and outcome for participant $i = 1, \dots, N_T$ in the trial. We obtain a point estimate and inference with TMLE as follows. A schematic flow chart in is provided in Figure~\ref{fig:tmle_flowchart}.

\begin{enumerate}
    
    \item Estimate the conditional expectation of the outcome given treatment assignment $A$ and covariates $W$, denoted $\mathbb{E}(Y | A, W)$. Precision is often improved by  using a ``working'' generalized linear model (GLM) adjusting for covariate(s) $W$ that are anticipated to be prognostic of the outcome \cite{rosenblum_simple_2010}. Precision is may further be improved by more flexibly estimating $\mathbb{E}(Y | A, W)$ with Super Learner using a diverse library of candidate estimators \cite{van_der_laan_super_2007}. Precision improvements are guaranteed when using Adaptive Pre-specification to select the optimal approach for estimating $\mathbb{E}(Y | A, W)$ \cite{balzer_adaptive_2016,balzer_adaptive_2024}.   
   
    \item Use the estimates from Step 1 to predict the outcome for all participants under intervention and control: $\hat{\mathbb{E}}(Y | A = 1, W_{i.})$ and $\hat{\mathbb{E}}(Y | A = 0, W_{i.})$ for $i = 1, \dots, N_T$.
    
    \item Estimate the propensity score, which is the probability of being randomized to the intervention given the covariates: $\mathbb{P}(A = 1 | W)$. Like Step 1, we could use a ``working'' GLM or Super Learner, but precision improvements are guaranteed when using Adaptive Pre-Specification \cite{balzer_adaptive_2016, balzer_adaptive_2024}. 
    
    \item Use the estimates from Step 3 to predict the propensity score for all participants: $\hat{\mathbb{P}}(A = 1 | W_{i.})$  for $i = 1, \dots, N_T$.
    
    \item Target the initial outcome predictions while simultaneously solving the efficient influence curve (IC) equation \cite{van_der_laan_targeted_2011}. Briefly, based on the efficient IC for the target effect,  we define a ``clever covariate'' that harnesses information in the estimated propensity score. Then we run a logistic outcome regression with the initial outcome predictions as offset and adjusting for the ``clever covariate''. With the resulting fluctuation parameters, we obtain targeted outcome predictions under the intervention and control $\hat{\mathbb{E}}^{*}(Y | A=1, W_{i.})$ and $\hat{\mathbb{E}}^{*}(Y | A=0, W_{i.})$ for  $i = 1, \dots, N_T$.
    \item Obtain a point estimate by averaging the targeted predictions under the intervention and the control
 $$\hat{\Psi}(1) = \frac{1}{N_T} \sum_{i=1}^{N_T} \hat{\mathbb{E}}^{*}(Y | A=1, W_{i.}) \text{ \quad and \quad } \hat{\Psi}(0)=  \frac{1}{N_T}  \sum_{i=1}^{N_T} \hat{\mathbb{E}}^{*}(Y | A=0, W_{i.})$$
    
    Then contrast on the scale of interest. Let $\hat{\Psi}$ be the resulting point estimate of the intervention effect. 

    \item Obtain a variance estimate $\hat{\sigma}^2$ with the sample variance of the estimated and aggregated influence curve, scaled by the number of independent units $J^*$ (Appendix~\ref{app:ICagg}).
\item With the point estimate $\hat{\Psi}$ and variance estimate $\hat{\sigma}^2$, create Wald-Type 95\% confidence intervals and test the null hypothesis.
\end{enumerate}

\section{Aggregation of influence curves}
\label{app:ICagg}

For notational convenience, let $m=1,\ldots, M$ index the independent units in the trial. For example, in a typical IRGTT, $M = K + L$, with $K$ the number of intervention arm clusters and $L$ the number of individuals who continue without clustering. By design, the TMLE $\hat{\Psi}$ for the individual-level effect $\Psi$ described in Appendix~\ref{app:tmle} solves the efficient influence curve equation: 
\begin{equation*}
    0 =  \frac{1}{N_T} \sum_{i=1}^{N_T} \hat{D}_{i.}
\end{equation*}
where $\hat{D}_{i.}$ is the estimated influence curve for participant $i=\{1,\ldots, N_T\}$. 
Following \cite{schnitzer_effect_2014, benitez_defining_2023}, we can re-express this in terms of the empirical mean of an aggregated influence curve accounting for the partially clustered design: 
\begin{equation*}
\begin{aligned}
0 &= \frac{1}{N_T} \sum_{m=1}^{M} \sum_{i=1}^{N_m} \hat{D}_{im} \\ 
&= \frac{1}{M} \sum_{m=1}^{M} \sum_{i=1}^{N_m} \frac{M}{N_T}\hat{D}_{im} \\\
   &= \frac{1}{M} \sum_{m=1}^{M} \hat{D}^*_m \text{\quad where \quad } \hat{D}^*_m =  \sum_{i=1}^{N_m} \frac{M}{N_T}\hat{D}_{im}
\end{aligned}
  \end{equation*}
Thus, the aggregated influence curve $\hat{D}^*_m$ is a weighted sum of the participant-level influence curves  in intervention clusters and a scaled version of the participant-level influence curve in control ``clusters''.
We emphasize that since the cluster size varies using a naive aggregation approach of  taking the empirical mean in each cluster (i.e., $1/N_m \sum_i \hat{D}_{im})$ is incorrect.

Under reasonable regularity conditions \cite{van_der_laan_targeted_2011, van_der_laan_targeted_2018}, 
 TMLE is asymptotically linear, meaning the estimator minus the effect behaves (up to a second-order term) as an empirical mean of the influence curve:
\begin{equation*}
\begin{aligned}
 \hat{\Psi} - \Psi
    &= \frac{1}{M} \sum_{m=1}^{M} D^*_m  \text{\quad where \quad } D^*_m =  \sum_{i=1}^{N_m} \frac{M}{N_T}D_{im}
\end{aligned}
  \end{equation*}
This provides the basis for inference through the Central Limit Theorem. Specifically, we estimate the variance of the TMLE  with the sample variance of the estimated and aggregated influence curve, scaled by the number of independent units:
   $$\hat{\sigma}^2=\frac{\hat{Var}(\hat{D}^*_m)}{M}$$
Implementation of this aggregation can have some counterintuitive impacts on variance estimation, as demonstrated in their application to SEARCH-IPT (Section~\ref{sipt_results}).

\clearpage
\newpage 

\section{Additional simulation results}
\label{app:moresims}

\subsection{Continuous outcomes, different data-generating processes}
\label{app:moresims_cont}

\begin{figure}[h]
    \centering
    \includegraphics[width = \textwidth]{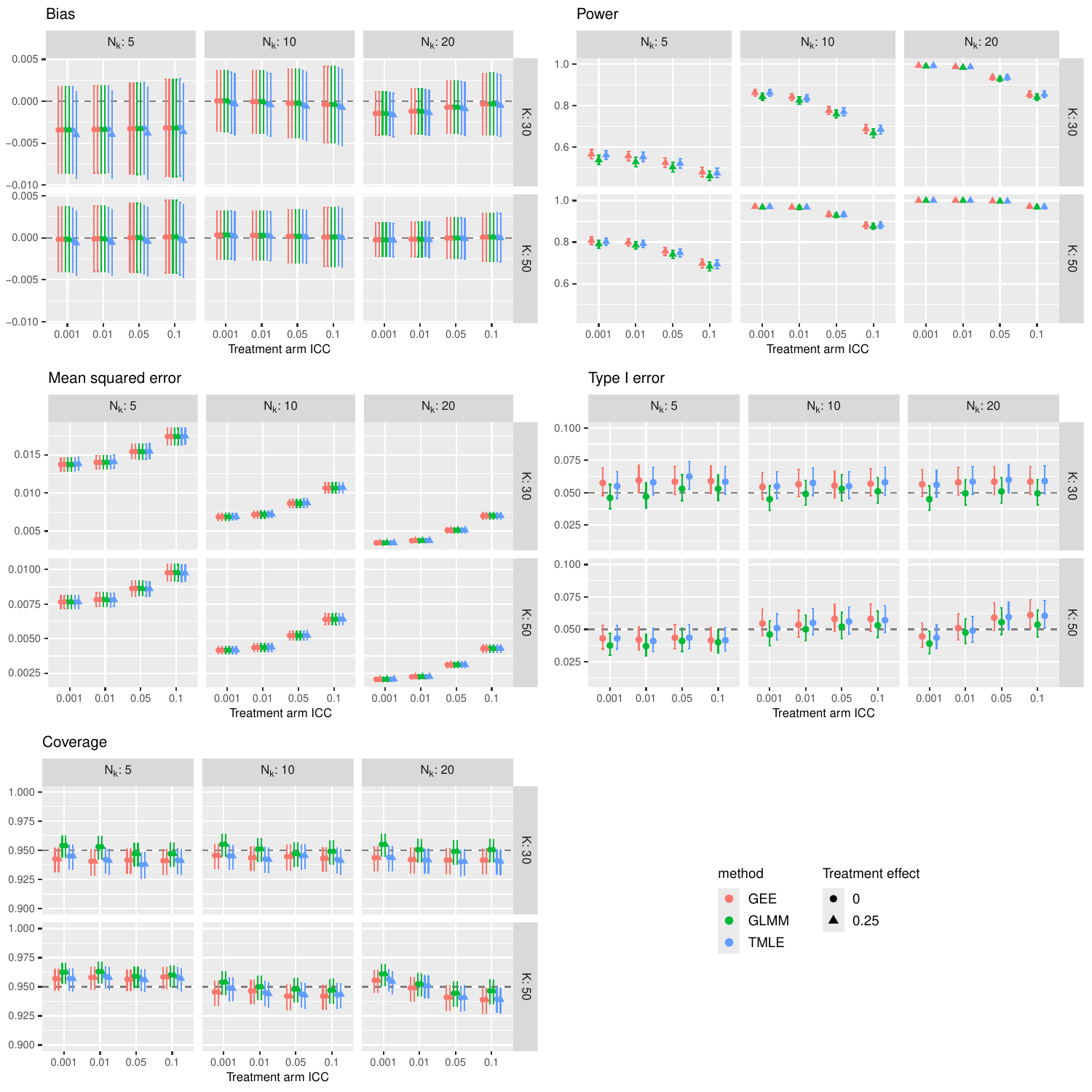}
    \caption{Simulations results for the ``Main terms" scenario with a continuous outcome.}
    \label{fig:main_terms}
\end{figure}

\begin{figure}
    \centering
    \includegraphics[width = \textwidth]{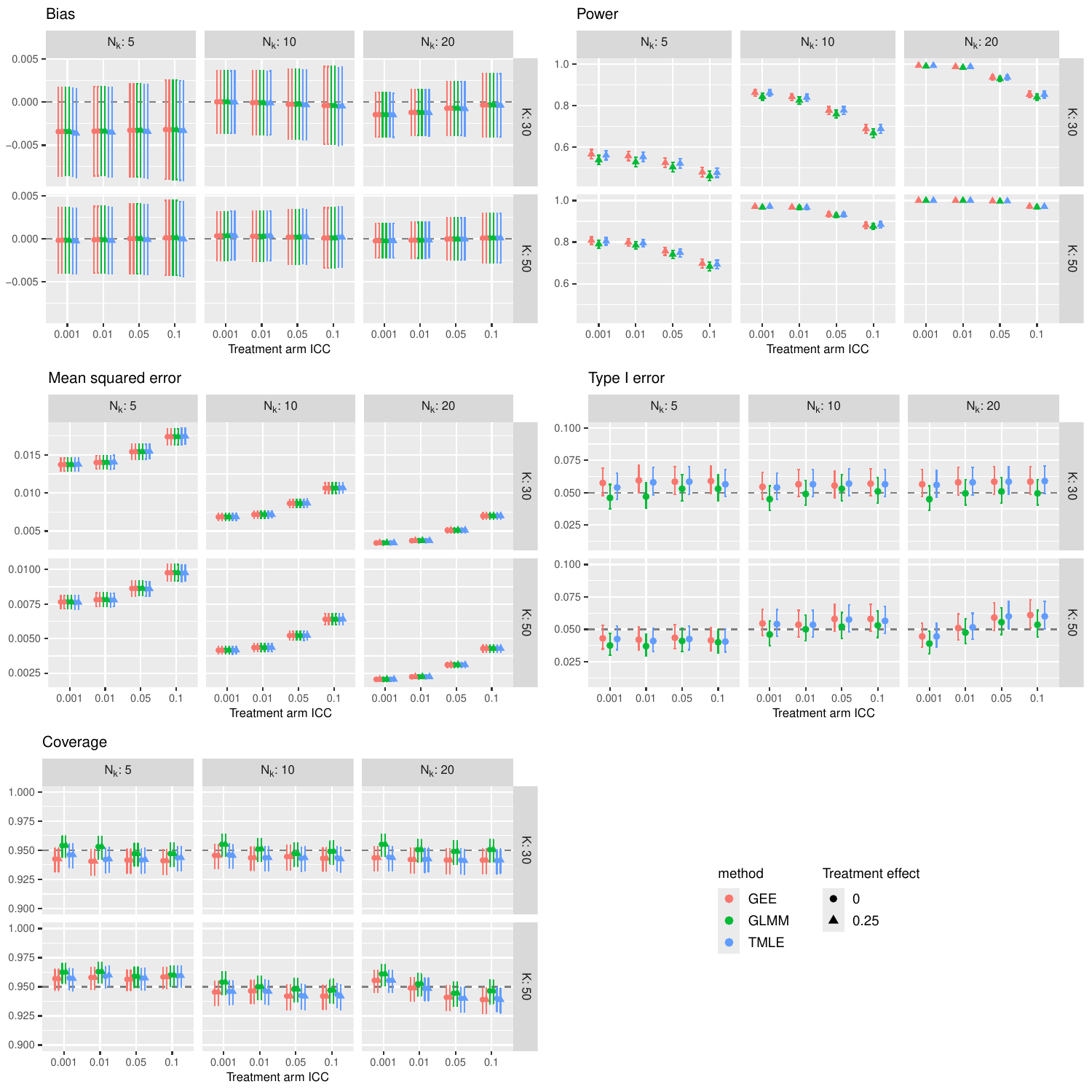}
    \caption{Simulations results for the ``Treatment only" scenario with a continuous outcome.}
    \label{fig:overfit}
\end{figure}

Figures~\ref{fig:main_terms} and \ref{fig:overfit} show the simulation results for a continuous outcome in the ``Main terms" and ``Treatment only" setting. As noted in the main text, all approaches showed similar coverage, power, mean squared error, bias, and maintained acceptable Type I error control.

\clearpage
\newpage

\subsection{Binary outcomes}
\label{app:moresims_bin}

Additional simulations were conducted for binary outcomes. We maintained the same clustering structures and same data generating process for the covariates and treatment assignment. For the outcome generation, we explored the following settings
\begin{enumerate}
   \item \emph{Complex:} $Y_{ij} = \mathbb{I}\big[ \text{logit}^{-1} (-1 + \beta \cdot A_{ij} + W0_{ij}^2 - .5 \cdot W1_{ij} + W0^2_{ij} \cdot W1_{ij} + UE_j) > UY_{ij} \big]$
    \item \emph{Main terms:} $Y_{ij} = \mathbb{I}\big[\text{logit}^{-1} ( -.5 + \beta \cdot A_{ij} + W0_{ij} + .5 \cdot W1_{ij} + UE_j) > UY_{ij} \big]$
    \item \emph{Treatment only:} $Y_{ij} = \mathbb{I}\big[\text{logit}^{-1} ( -2 + \beta \cdot A_{ij} + UE_j) > UY_{ij} \big]$
\end{enumerate}
where $UY_{ij} \sim \text{Uniform}(0,1)$ and $\beta \in \{0,.35\}$. For estimation and inference, GLMM and GEE were implemented as before, but with the logistic link function. The implementation of TMLE did not change. Due to the variety of ways to parameterize an ICC for binary outcomes \cite{wu_comparison_2012}, the treatment arm ICCs vary slightly from the continuous case.

Given these implementations and the non-collapsibility of odds ratios, the 3 estimators targeted different effects. The GLMM estimated a conditional odds ratio --- conditional on the covariates and cluster membership. The GEE estimated a conditional odds ratio --- conditional on the covariates. TMLE estimated a marginal odds ratio, after adaptively adjusting for the covariates and then standardizing them to the covariate distribution. It is worth noting that TMLE also estimated the risk difference and risk ratio, but we focused on the odds ratio to improve comparability across approaches. Given the different targets of inference, we simplified our presentation to  the Type I error rate (for $\beta = 0$), attained power (for $\beta \neq 0$), and the ratio of the estimated standard error to the true standard deviation of the point estimates (``SE/SD ratio") across the 2000 simulated data sets.

\begin{figure}[h]
    \centering
    \includegraphics[width = \textwidth]{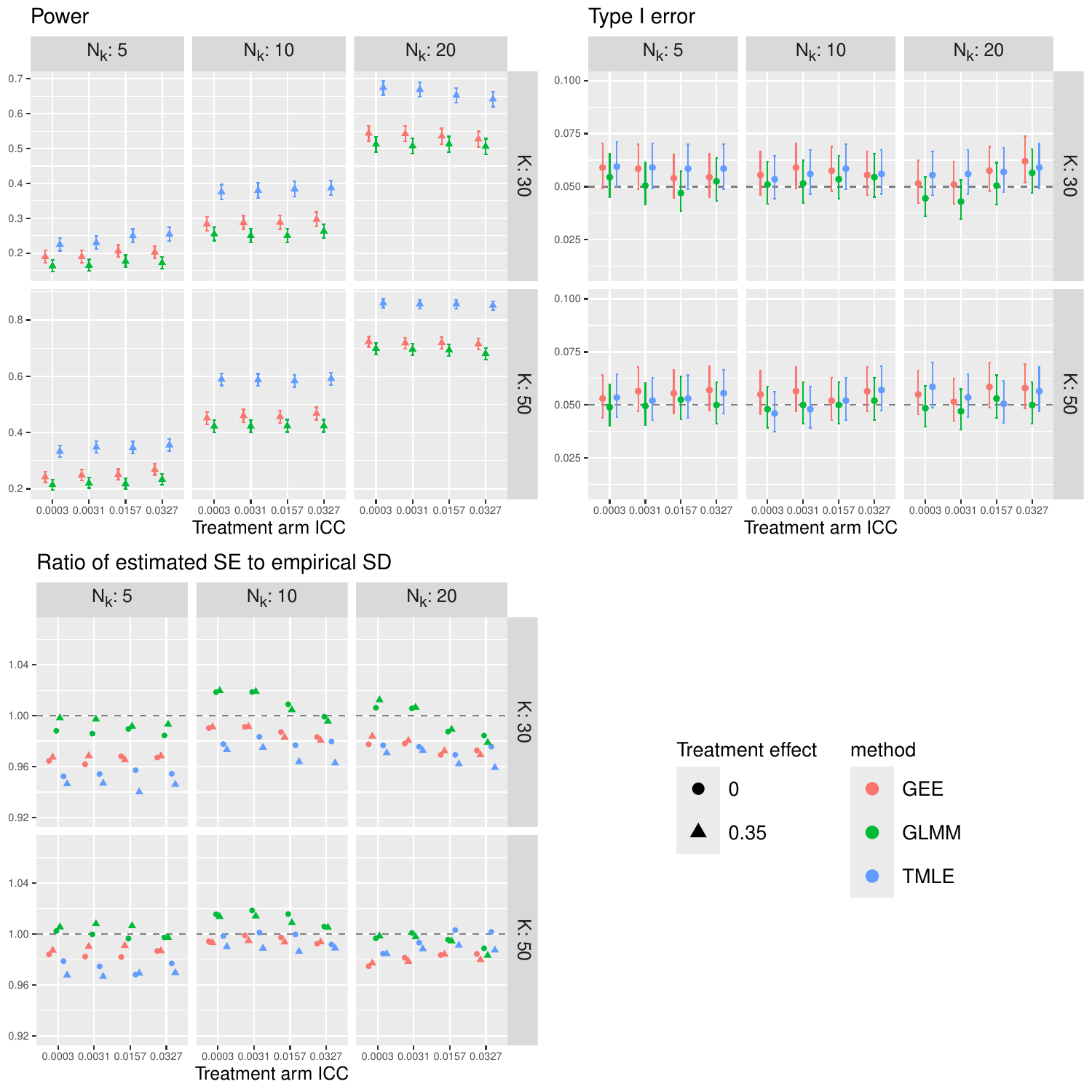}
    \caption{Simulation results for the ``Complex" scenario with a binary outcome.}
    \label{fig:complex_covars_binary}
\end{figure}

\begin{figure}
    \centering
    \includegraphics[width = \textwidth]{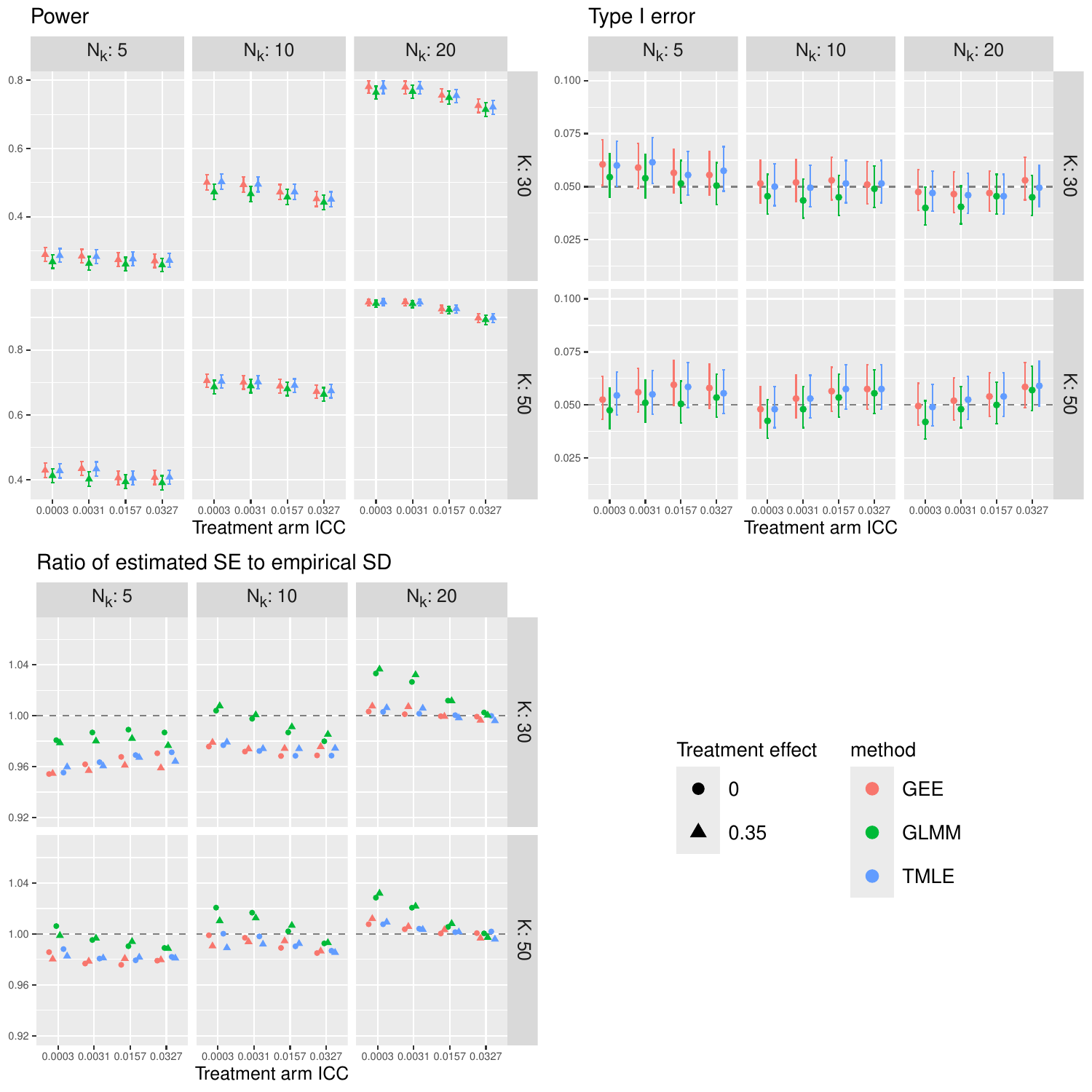}
    \caption{Simulation results for the ``Main terms''  scenario with a binary outcome.}
    \label{fig:main_terms_binary}
\end{figure}

\begin{figure}
    \centering
    \includegraphics[width = \textwidth]{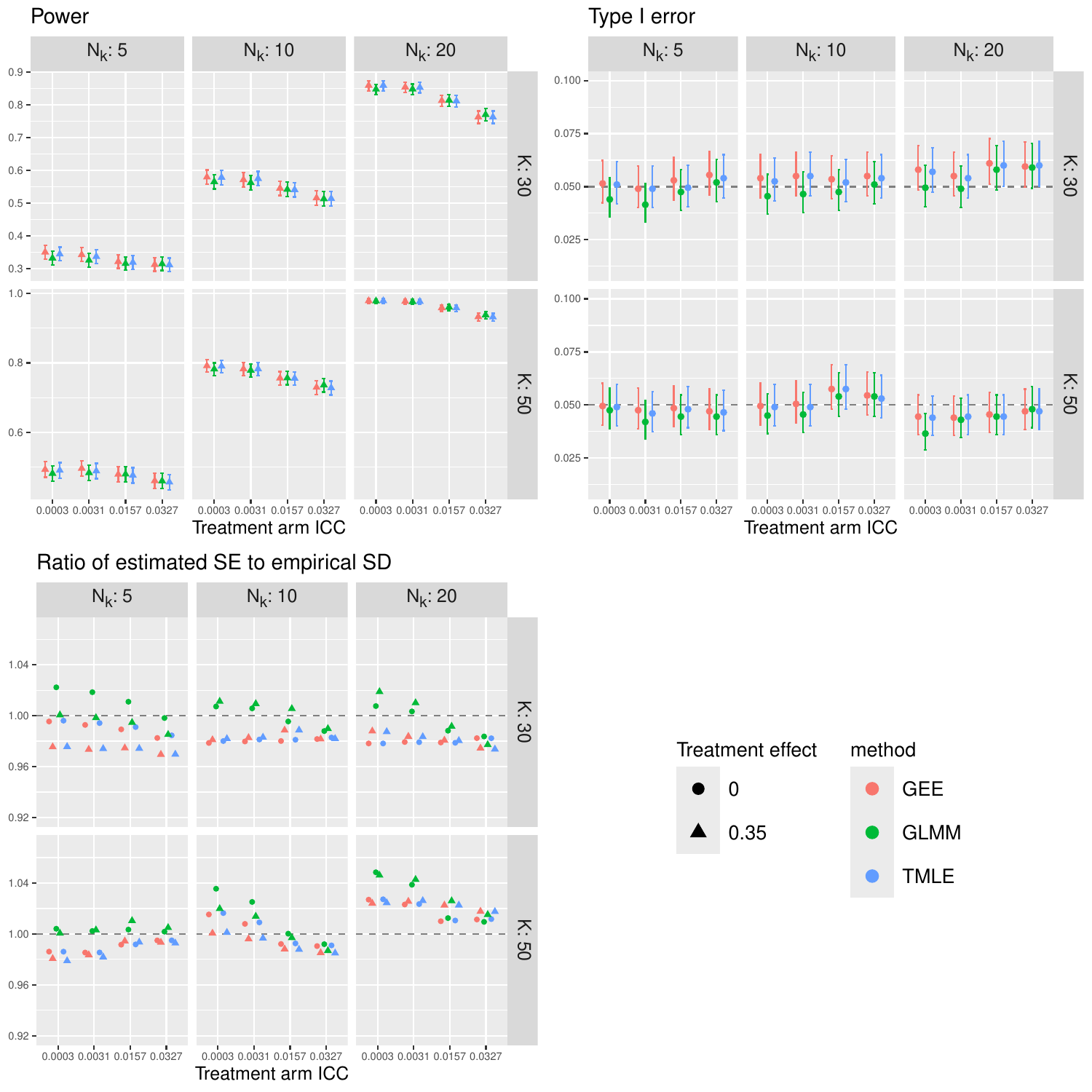}
    \caption{Simulation results for the ``Treatment only" scenario with a binary outcome.}
    \label{fig:overfit_binary}
\end{figure}

The overall results were similar to those for continuous outcomes. In the ``Complex'' scenario (Figure~\ref{fig:complex_covars_binary}), TMLE was equivalent or superior to GEEs and GLMMs on all measures. Under most parameter combinations, the attained power was notably higher for TMLE, while Type I error control was maintained. Figures~\ref{fig:main_terms_binary} and \ref{fig:overfit_binary} show the results under the ``Main terms'' and ``Treatment only'' scenarios, where power, Type I error control, and SE/SD ratio were similar across estimation approaches.

\clearpage
\newpage

\subsection{Continuous outcomes with an uneven numbers of participants per arm}
\label{app:moresims_cont_unbalanced}

\begin{figure}[h]
    \centering
    \includegraphics[width = \textwidth]{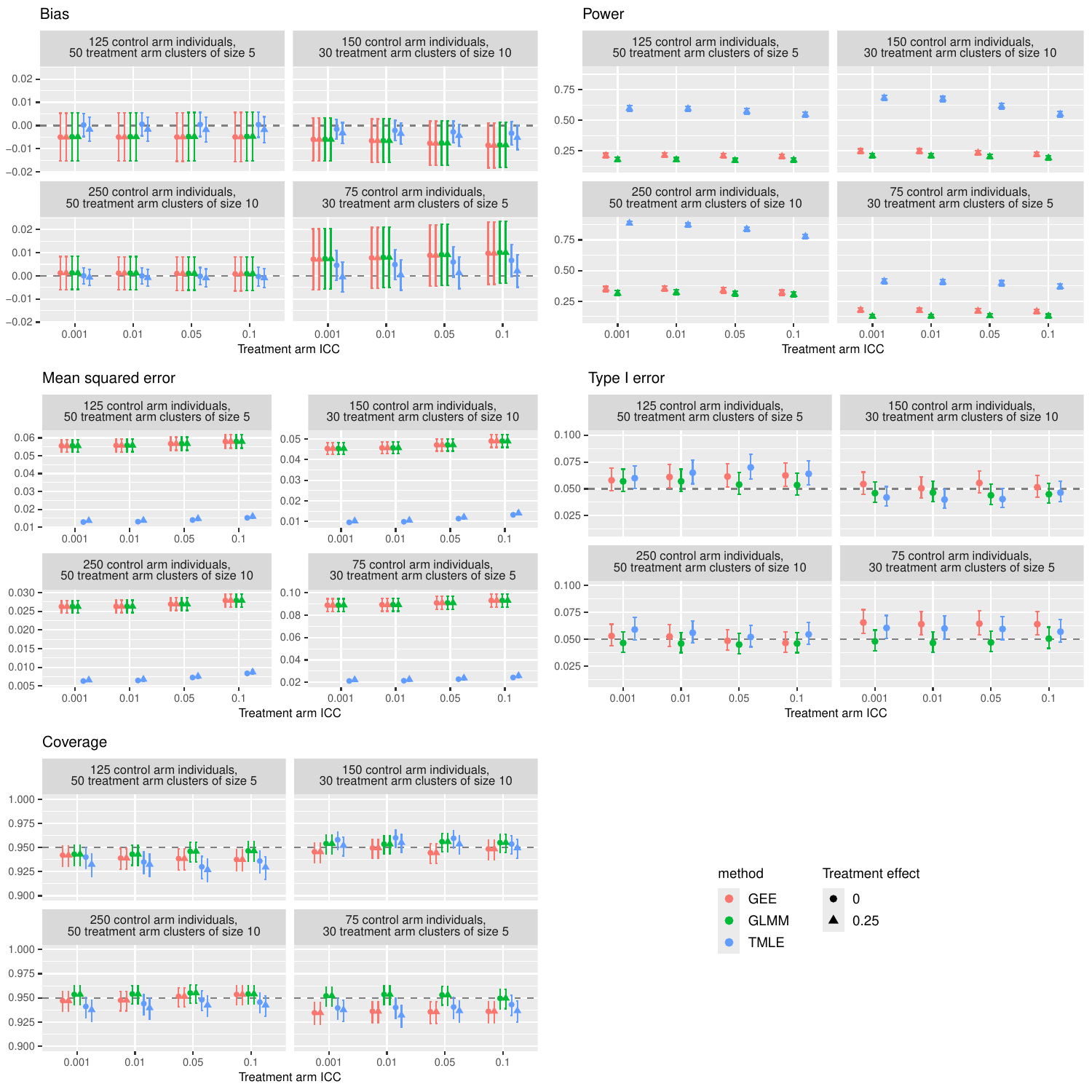}
    \caption{Simulations results for the ``Complex" scenario with a continuous outcome and uneven numbers of participants per arm.}
    \label{fig:irgtt_complex_continuous_ub}
\end{figure}

\begin{figure}
    \centering
    \includegraphics[width = \textwidth]{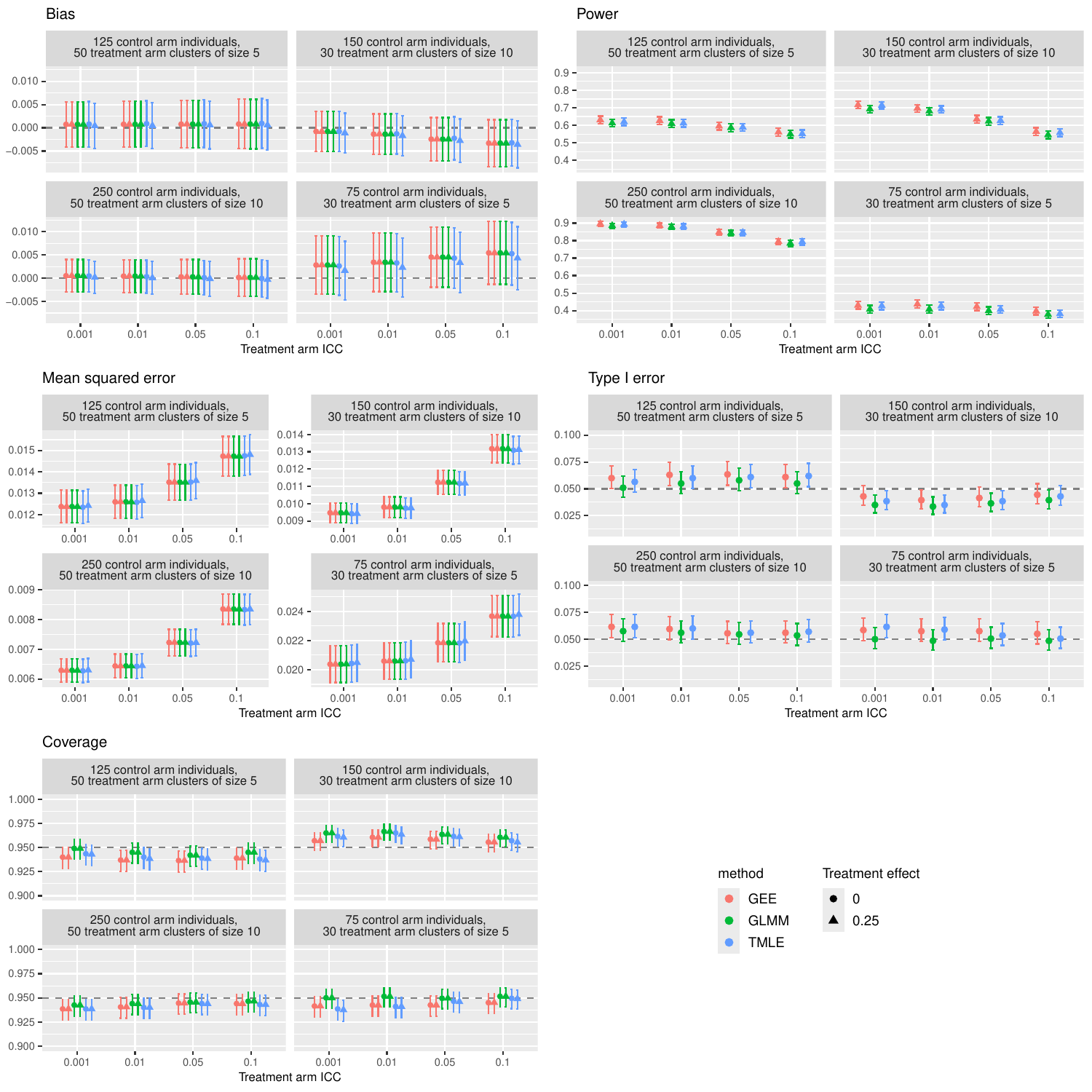}
    \caption{Simulations results for the ``Main terms" scenario with a continuous outcome and uneven numbers of participants per arm.}
    \label{fig:irgtt_main_terms_continuous_ub}
\end{figure}

\begin{figure}
    \centering
    \includegraphics[width = \textwidth]{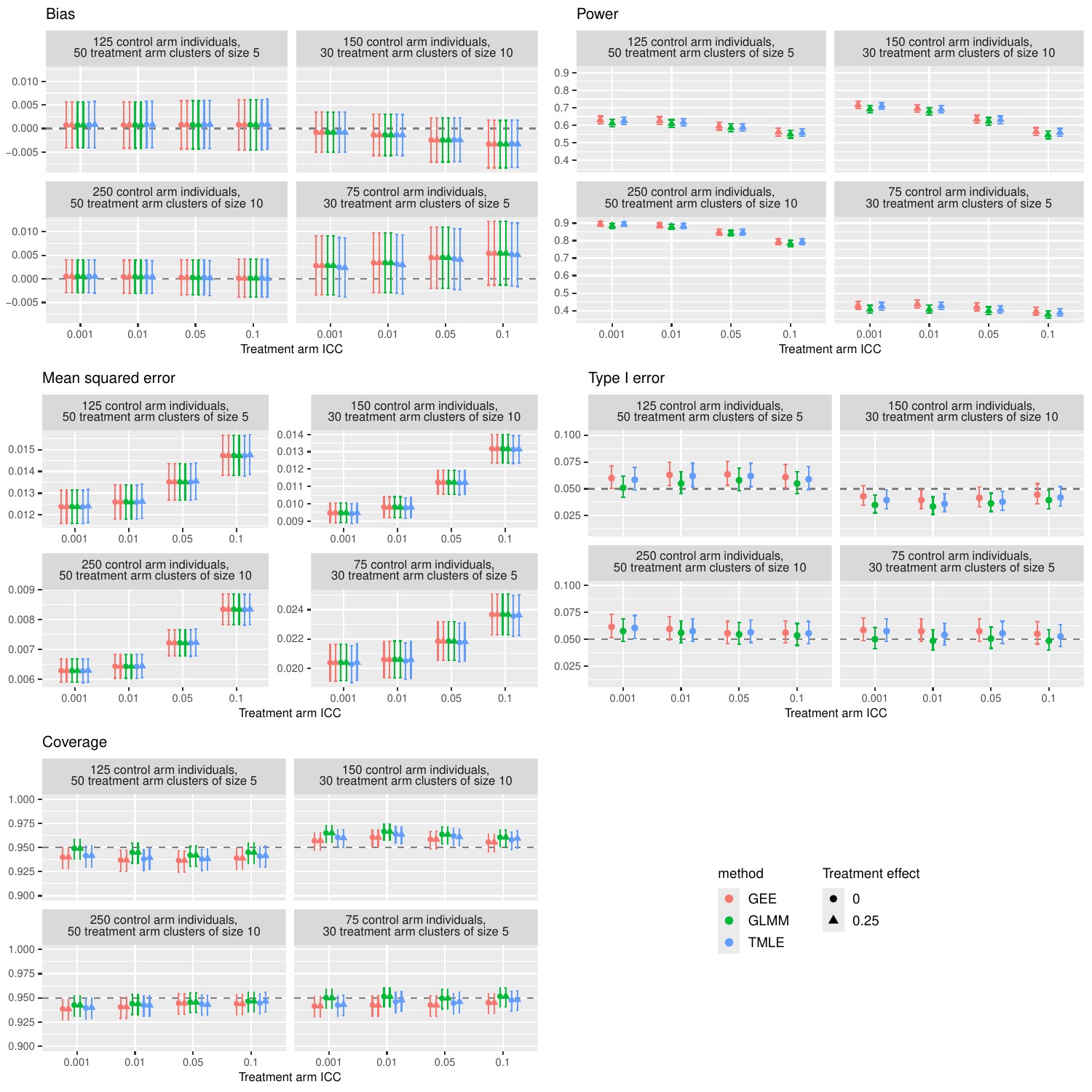}
    \caption{Simulations results for the ``Treatment only" scenario with a continuous outcome and uneven numbers of participants per arm.}
    \label{fig:irgtt_overfit_continuous_ub}
\end{figure}

\clearpage
\newpage

\subsection{Binary outcomes with an uneven numbers of participants per arm}
\label{app:moresims_bin_unbalanced}

\begin{figure}[h]
    \centering
    \includegraphics[width = \textwidth]{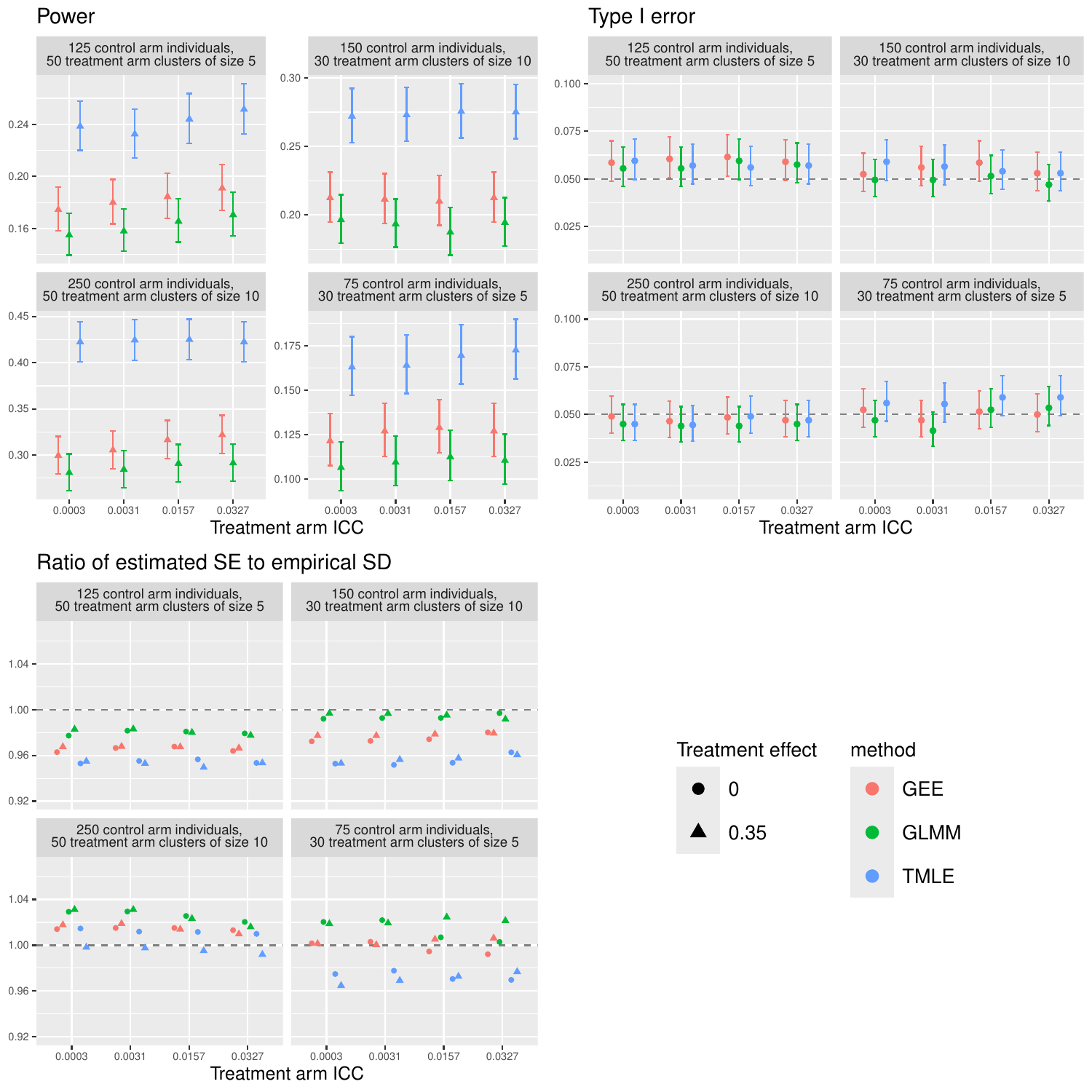}
    \caption{Simulations results for the ``Complex" scenario with a binary outcome and uneven numbers of participants per arm.}
    \label{fig:irgtt_complex_binary_ub}
\end{figure}

\begin{figure}
    \centering
    \includegraphics[width = \textwidth]{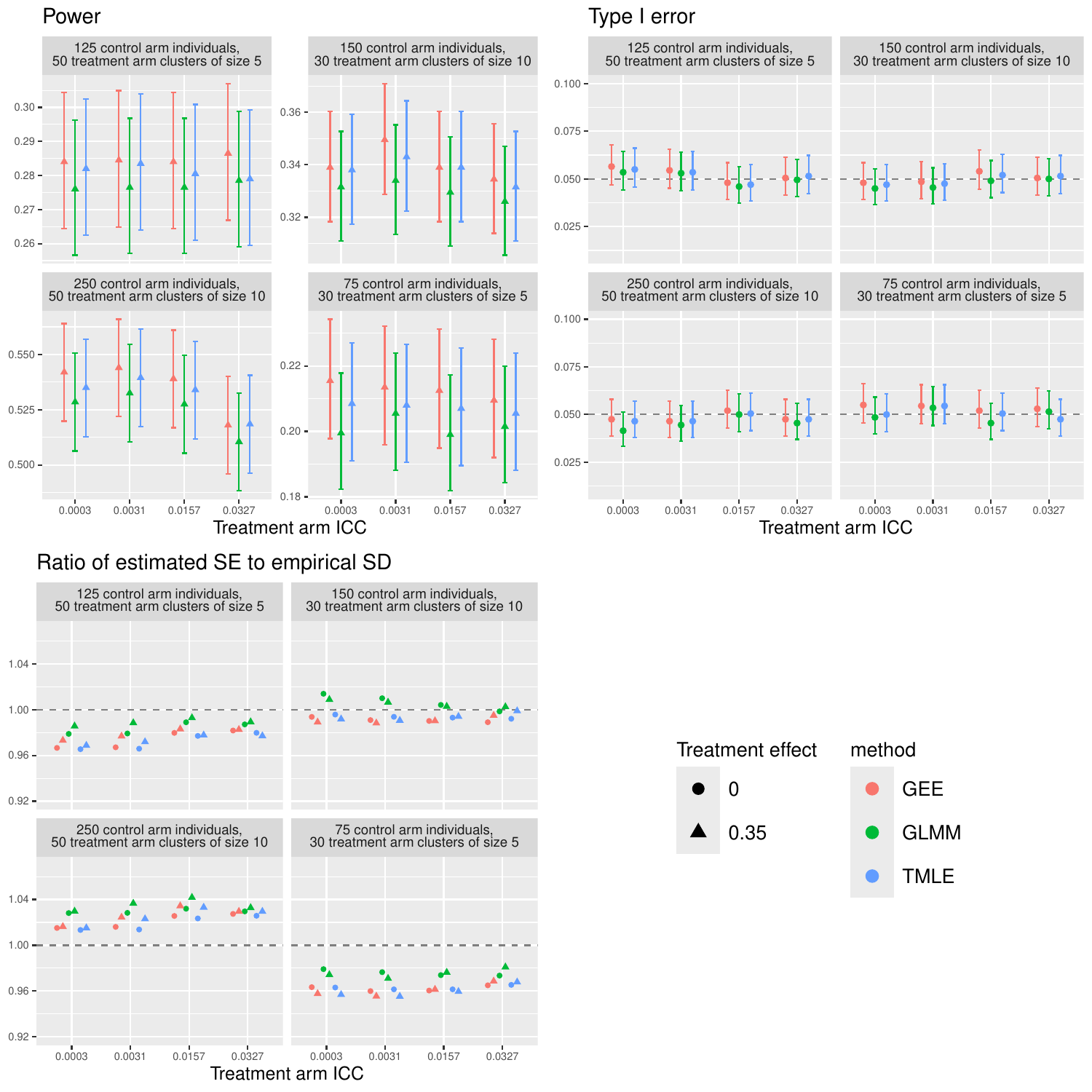}
    \caption{Simulations results for the ``Main terms" scenario with a binary outcome and uneven numbers of participants per arm.}
    \label{fig:irgtt_main_terms_binary_ub}
\end{figure}

\begin{figure}
    \centering
    \includegraphics[width = \textwidth]{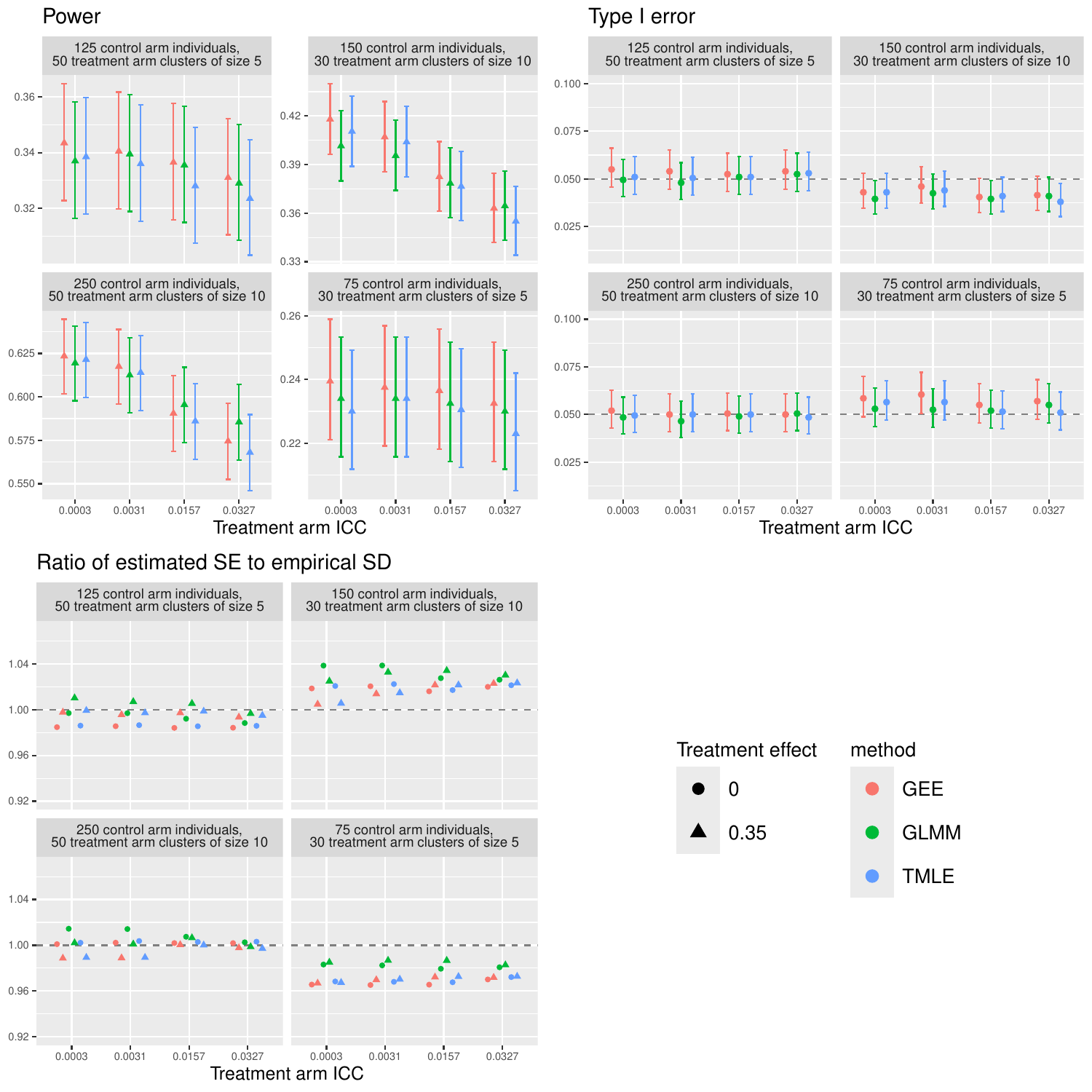}
    \caption{Simulations results for the ``Treatment only" scenario with a binary outcome and uneven numbers of participants per arm.}
    \label{fig:irgtt_overfit_binary_ub}
\end{figure}

\clearpage
\newpage

\subsection{Continuous outcomes, with and without IC aggregation}
\label{app:sims_no_agg}

\begin{figure}[h]
    \centering
    \includegraphics[width = \textwidth]{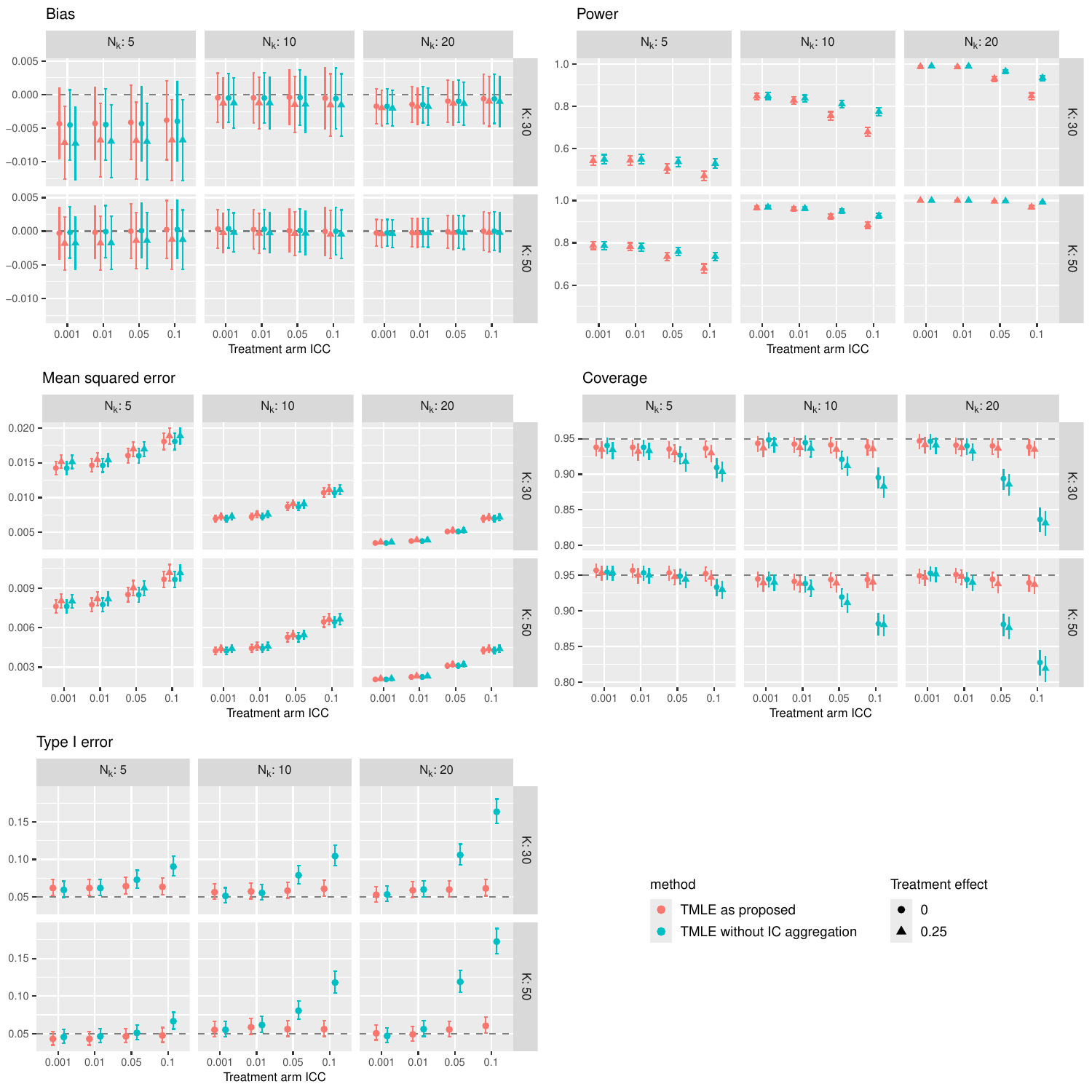}
    \caption{Simulations results for the ``Complex" scenario with a continuous outcome, comparing proposed method using TMLE influence curve aggregation vs. TMLE without influence curve aggregation (thus assuming independence among all units).}
    \label{fig:irgtt_complex_ms}
\end{figure}


\end{document}